\documentclass[a4paper,fleqn]{cas-sc}

\usepackage[square,numbers]{natbib}
\usepackage{graphicx}
\usepackage[normalem]{ulem}
\useunder{\uline}{\ul}{}
\usepackage{breakurl}

\usepackage{booktabs}
\usepackage{eurosym}
\usepackage{breakurl}
\usepackage{todonotes}
\usepackage[normalem]{ulem}

\usepackage{hyperref}
\usepackage{color}
\usepackage{amssymb}
\usepackage{rotating}
\usepackage[T1]{fontenc}
\usepackage{lscape}
\usepackage[autostyle]{csquotes}
\usepackage[inline]{enumitem}
\usepackage[ruled,vlined]{algorithm2e}

\usepackage[utf8]{inputenc}
\usepackage{enumerate}
\usepackage{mathtools}
\usepackage{microtype}
\usepackage{longtable}
\usepackage{pdflscape}
\usepackage{graphicx}
\usepackage{tabularx}
\usepackage{multirow}
\usepackage{textcomp}
\usepackage{csquotes}
\usepackage{booktabs}
\usepackage{graphics}
\usepackage{xr-hyper}
\usepackage{hyperref}
\usepackage{amsmath}
\usepackage{amssymb}
\usepackage{natbib}
\usepackage{epsfig}
\usepackage{lscape}
\usepackage{lineno}
\usepackage{array}
\usepackage{float}
\usepackage{todonotes}
\usepackage{tabu}
\usepackage{siunitx}

\usepackage{acronym}
\usepackage{multicol}

\def\tsc#1{\csdef{#1}{\textsc{\lowercase{#1}}\xspace}}
\tsc{WGM}
\tsc{QE}

\begin{document}
\let\WriteBookmarks\relax
\def\floatpagepagefraction{1}
\def\textpagefraction{.001}

\shorttitle{Insights and Challenges in Deploying ChatGPT and Generative Chatbots for FAQs}    

\shortauthors{F. Khennouche et al.}  

\title [mode = title]{Revolutionizing Customer Interactions: Insights and Challenges in Deploying ChatGPT and Generative Chatbots for FAQs}  

\author[1]{Feriel Khennouche}
\ead{khennouche@estin.dz}
\author[1,2]{Youssef Elmir}
\ead{elmir@estin.dz}

\author[3]{Nabil Djebari}[]
\ead{djebari@estin.dz}
\author[4]{Yassine Himeur}\cormark[1]
\ead{yhimeur@ud.ac.ae}

\author[5,6]{Abbes Amira}[]
\ead{aamira@sharjah.ac.ae}

\address[1]{Laboratoire LITAN, École supérieure en Sciences et Technologies de l’Informatique et du Numérique, RN 75, Amizour 06300, Bejaia, Algeria}
\address[2]{SGRE-Lab, University Tahri Mohammed of Bechar, Bechar 08000, Bechar, Algeria}

\address[3]{Laboratoire LIMED, Faculté des Sciences Exactes, Université de Bejaia, , Bejaia 06000, Algeria}

\address[4]{College of Engineering and Information Technology, University of Dubai, Dubai, UAE}

\address[5]{Department of Computer Science, University of Sharjah, , Sharjah , UAE}

\address[6]{Institute of Artificial Intelligence, De Montfort University, , Leicester , United Kingdom}

\cortext[1]{Corresponding author}



\begin{abstract}
In the rapidly evolving domain of artificial intelligence, chatbots have emerged as a potent tool for various applications ranging from e-commerce to healthcare. This research delves into the intricacies of chatbot technology, from its foundational concepts to advanced generative models like ChatGPT. We present a comprehensive taxonomy of existing chatbot approaches, distinguishing between rule-based, retrieval-based, generative, and hybrid models. A specific emphasis is placed on ChatGPT, elucidating its merits for frequently asked questions (FAQs)-based chatbots, coupled with an exploration of associated Natural Language Processing (NLP) techniques such as named entity recognition, intent classification, and sentiment analysis. The paper further delves into the customization and fine-tuning of ChatGPT, its integration with knowledge bases, and the consequent challenges and ethical considerations that arise. Through real-world applications in domains such as online shopping, healthcare, and education, we underscore the transformative potential of chatbots. However, we also spotlight open challenges and suggest future research directions, emphasizing the need for optimizing conversational flow, advancing dialogue mechanics, improving domain adaptability, and enhancing ethical considerations. The research culminates in a call for further exploration in ensuring transparent, ethical, and user-centric chatbot systems.
\end{abstract}




\maketitle

\section{Introduction}
Large Language Models (LLMs) and generative AI have become vital components of human-computer interaction in recent years. These systems offer unprecedented potential to bridge the communication gap between humans and machines by allowing computers to understand and generate human-like text \cite{kheddar2023deep,sohail2023decoding}. LLMs like GPT-4 can generate contextually relevant, coherent, and complex responses, moving beyond basic keyword recognition to a more holistic understanding of human language \cite{farhat2023analyzing}. They enable a natural language interface, thereby democratizing access to technology, especially for those who may not be technically skilled. In an increasingly digital age, the ability of generative AI to simplify complex information, answer queries, create content, and even predict user needs is proving to be invaluable \cite{sohail2023future}. It is enabling a new generation of applications and services, from virtual assistants to content generation, personalized education, and more. The importance of LLMs and generative AI in human-computer interaction can not be overstated; they are shaping the future of how we communicate, interact, and coexist with technology \cite{sohail2023using}.

\begin{figure}[!h]
\centering
\includegraphics[width=0.9\textwidth]{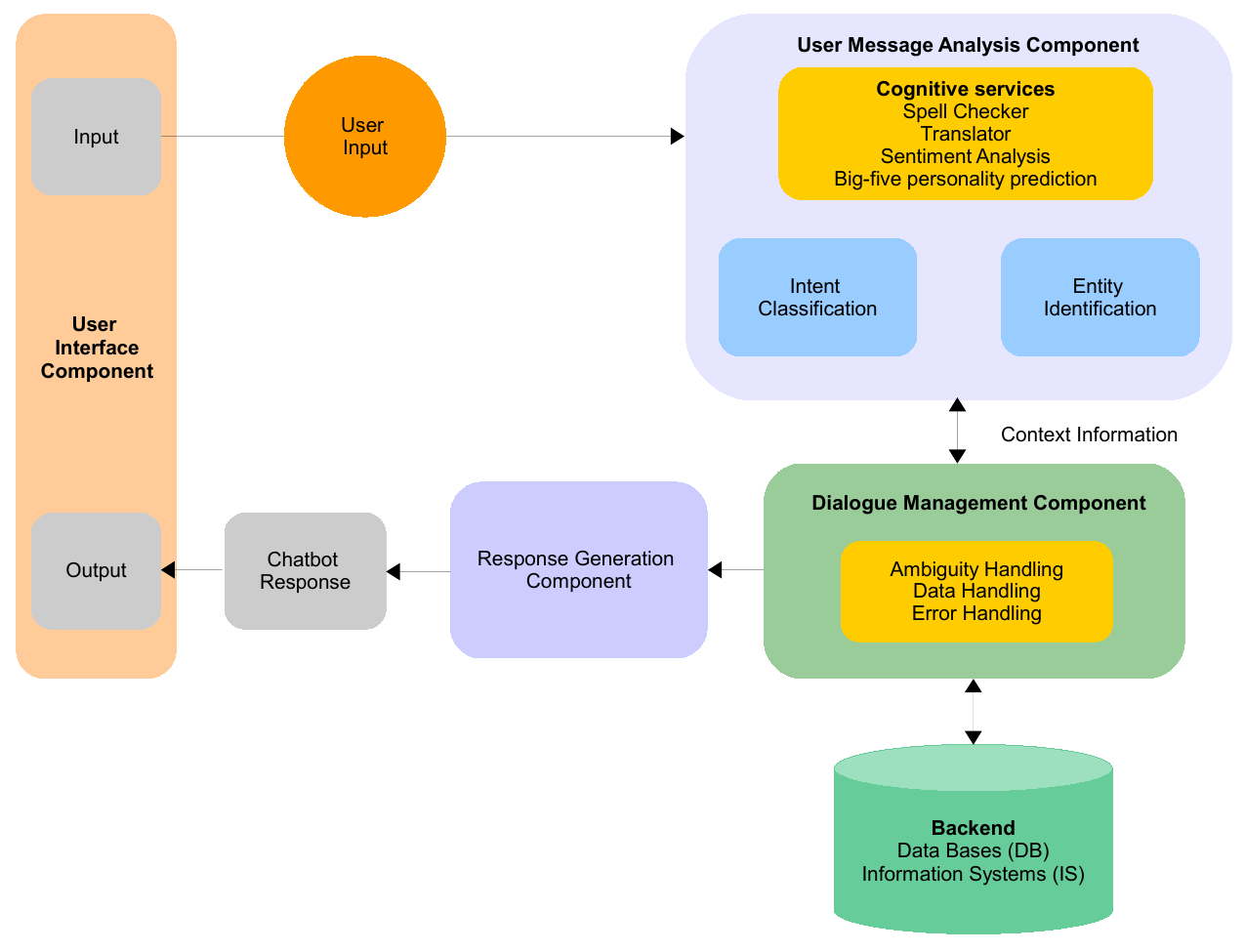}
\caption{\textcolor{black}{General chatbot structure.}\label{figchat}}
\end{figure}

In addition to Google's models like GPT and OpenAI's GPT-3, there are several other remarkable generative tools in the field of AI-driven content creation. Projects like "Bard" aspire to compose poetry and literature, showcasing AI's creative potential \cite{waisberg2023google}. Furthermore, Explainable AI (XAI) frameworks are gaining prominence, managing the challenge of making complex AI systems more interpretable and transparent. These tools, along with others, are collectively contributing to advancements in diverse domains, while shared challenges such as ethical considerations, bias mitigation, and fine-tuning the balance between creativity and control continue to shape the terrain of AI-generated content. A chat dialogue system, chatbot, or conversational agent is a computer program that can hold a natural language conversation. A chatbot is able to respond user's request automatically \cite{lin2023review}. Its main goal is to enhance online support by using a frequently asked question (FAQ)-based chatbot to provide quick and precise answers to user queries, assisting with tasks and inquiries. \cite{jesus2023conversational}. The biggest challenge in implementing a FAQ chatbot is to keep the human aspect in communication, as it will be degraded when communicating with a bot. Typically, \cite{sethi2020faq} recommends keeping these facts when implementing a FAQ chatbot: (i) give it a Personality (the dialogue style should follow the business orientation),(ii) let it tie your Brand (the chatbot should be representative of your brand image), (iii) make it human-friendly, (iv) keep the conversation simple, (v) fewer typing, more clicking, and (vi) picking bot format (Rule-based ChatBot, Natural Language Processing (NLP)-based Bot, or Bot that uses a combination of the two). Figure. \ref{figchat} illustrates the basic chatbot architecture, where a user asks a question, and the chatbot responds based on its model and data. Details can vary based on the specific chatbot approach.

OpenAI \footnote{https://openai.com/}, a company specializing in artificial reasoning, also released at the end of 2022 a cutting-edge technology, in the field of NLP, a chatbot, baptized ChatGPT \cite{biswas2023role}. This chatbot uses deep learning algorithms to generate human-like responses to text-based inputs, and it is widely used in various applications such as language translation, question answering, and chatbot development \cite{lund2023chatting}. It has proven to be very effective in improving the accuracy and efficiency of these applications, making them more user-friendly and accessible. Hence, this chatbot has the potential to significantly enhance the speed and efficiency of scientific research, ultimately leading to new discoveries and breakthroughs in various fields \cite{king2023conversation}. The latest research studies have demonstrated that implementing a chatbot with a FAQ-based system can enhance user experience and accuracy compared to traditional FAQ pages \cite{zhao2023comparing}. Consequently, this paper explores the latest AI technologies for FAQ-based chatbots, reviews existing models, and assesses their functions to enhance understanding of chatbot operations. It also examines the potential of using ChatGPT for FAQs. The paper addresses various research questions regarding state-of-the-art chatbot and ChatGPT technologies for FAQs, highlighting their strengths, limitations, and future research areas. Typically, the following research questions have been answered:


\begin{itemize}

\item What are the characteristics of a high-performing chatbot?
\item Can ChatGPT be utilized to improve chatbot performance?
\item What are the limitations of a chatbot? How can we handle ChatGPT limitations (ethical issues, security, etc.)?
\item How can I evaluate the effectiveness of a chatbot?
\item What are the different types of chatbots used for handling FAQs, and what are their key features and functionalities?
\item What are the existing methodologies and techniques employed in the development and implementation of chatbots and ChatGPT for FAQs?
\item What are the challenges and limitations associated with chatbots and ChatGPT for FAQs, both from a technical and user perspective?
\item How effective are chatbots and ChatGPT in providing accurate and relevant answers to frequently asked questions?
\item What are the best practices and strategies for designing and deploying chatbots and ChatGPT for FAQ purposes?
\item How do chatbots and ChatGPT impact user experience, engagement, and satisfaction when interacting with FAQ-based systems?
\item In the context of managing uncertainty, rectifying errors, and fostering collaboration between AI and humans, how do chatbots effectively combine recommender systems and explainable AI techniques?
\item What are the current trends and future directions in the development and improvement of chatbots and ChatGPT for FAQs?

\end{itemize}

\begin{center}
\begin{tabular}{|m{16cm}|}
\hline
{\small \textbf{Abbreviations:}} \\ 
\begin{multicols}{2}
\footnotesize

\begin{acronym}[AWGN] 
\acro{AI}{artificial intelligence}
\acro{AIML}{The Artificial Intelligence Markup Language}
\acro{BLEU}{BiLingual Evaluation Understudy}
\acro{BiGRU}{BiDirectional Gated Recurrent Unit}
\acro{BoW}{Bag-of-Words}
\acro{ChatGPT}{Chat Generative Pre-trained Transformer} 
\acro{CGC-QG}{Clue Guided Copy Network for Question Generation}
\acro{DQNs}{Deep Q Networks}
\acro{FAQ}{Frequently Asked Questions}
\acro{GRU}{Gated Recurrent Unit}
\acro{LLMs}{Large language models}
\acro{LSA}{Latent Semantic Analysis}
\acro{LSTM}{Long Short-Term Memory}
\acro{METEOR}{Metric for Evaluation for Translation with Explicit Ordering}
\acro{MILA}{Montreal Institute for Learning Algorithms}
\acro{NER}{Named Entity Recognition}
\acro{NLG}{Natural Language Generation}
\acro{NLP}{Natural Language Processing}
\acro{POS}{Part-of-Speech}
\acro{RNN}{Recurrent Neural Network}
\acro{ROUGE}{Recall Oriented Understudy for Gisting Evaluation}
\acro{SSA}{Sensibleness and Specificity Average}
\acro{seq2seq}{sequence-to-sequence}

\end{acronym}

\end{multicols} \\
\hline
\end{tabular}
\end{center}

\subsection{Paper Contributions}
This paper outlines a comprehensive research approach that begins by introducing the methodology employed, which involves an extensive review of various research databases. The study then proceeds to explore and discuss different evaluation metrics specifically designed for assessing the performance of chatbot systems. The subsequent section provides a comprehensive background on chatbots, setting the stage for the research scope. To categorize and analyze chatbot models effectively, a taxonomy is introduced, which classifies them into four distinct categories: Rule-based chatbots, Retrieval-based chatbots, Generative chatbots, and Hybrid chatbots. This taxonomy serves as a framework for understanding the landscape of chatbot models. The central focus of this paper is on ChatGPT, and in the main section, an overview of ChatGPT is provided. The decision to place ChatGPT at the core of this paper comes from its key pertinence within the domain of NLP and artificial intelligence. As an advanced conversational AI model, ChatGPT has not only gained significant progress but has also grown pertinent debates about ethics, biases, and its broader societal impact. By focusing on ChatGPT, this paper aims to thoroughly explore its applications, delve into potential controversies, and offer a comprehensive analysis of its implications, thereby contributing to a deeper understanding of this innovative technology's capabilities and limitations. In this paper we: 

\begin{itemize}
    \item Expose the advantages of ChatGPT in the context of FAQ chatbots.
    \item Highlight how ChatGPT leverages NLP techniques.
    \item Delve into the training and fine-tuning of FAQ chatbots using ChatGPT.
    \item Explore the potential for customization in ChatGPT, enabling the creation of unique and branded FAQ chatbot experiences.
    \item Take a comprehensive look at the limitations and ethical considerations associated with ChatGPT usage to ensure a well-rounded perspective.
    \item Present a range of practical applications for ChatGPT, showcasing its versatility and real-world utility across various domains.
    \item Explore the general challenges encountered during the development of any chatbot system, shedding light on the complexities developers face and offering insights into potential solutions.
\end{itemize}
The paper emphasizes the need for innovation in chatbot technology to overcome limitations and maximize its potential. It contributes to understanding ChatGPT, its applications, challenges, and future enhancements for chatbot systems. Unlike the existing state of the art papers on chatbots, this paper aims to make a significant contribution by creating a taxonomy of existing FAQ chatbot models and situating ChatGPT, one of the most well-known chatbots, among them. Taxonomies have proven to be useful in various fields to categorize and classify complex concepts or notions. In the domain of chatbots, taxonomies have been used to structure and index the various types and functions of chatbots. In this paper, The main contribution of this paper is the development of a taxonomy of existing FAQ chatbot models. Through the classification and analogizing of different models. Thus, the aim is to provide a more useful understanding of their strengths and weaknesses and to identify areas for advancement. This taxonomy not only serves as a framework for this study but also provides a valuable resource for future research and development of FAQ chatbots. The proposed taxonomy is based on a thorough review of the literature and consultation with experts in the field. It is believed to contribute significantly to the advancement of knowledge and understanding of FAQ chatbots and their potential applications.

\subsection{Review Methodology}
The present review study is based on the protocols and procedures suggested in \cite{kitchenham2004procedures}, which has recently been adopted widely in many studies such as \cite{himeur2021survey}. During the rigorous literature review process, a comprehensive array of reputable databases, including Scopus, Clarivate, IEEExplore, ScienceDirect, and Springer, was meticulously utilized to source relevant articles. The study's focal point was notably oriented towards the realm of chatbots, with a specific emphasis on the intricacies of FAQ-based chatbots. The overarching objective was to meticulously scrutinize and assess the existing methodologies in this particular facet of the chatbot landscape. To ensure a well-defined scope, the search criteria were meticulously calibrated. The chosen timeframe encompassed the last decade, encompassing a selection of papers published within the past 10 years. A judicious exclusion of unpublished works was maintained to uphold the study's scholarly rigor. Strategic keyword deployment formed an integral part of the search strategy. Keywords such as "Chatbot," "FAQ chatbot," "Chatbot evaluation," "Chatbot model," and "Chatbot issues" were adeptly employed to effectively sift through the extensive literature and extract the most pertinent materials. The graphical representation above succinctly captures the trajectory of research in the chatbot FAQs domain over the past decade, utilizing data sourced from the comprehensive Google Scholar database. One striking trend stands out prominently – a significant surge in research activity over the last five years. This remarkable upswing can be attributed to the growing integration of artificial intelligence across multifarious sectors, encompassing realms like marketing, sales, education, health, and beyond. A main deep motivation driving this research is the compelling need to address the mounting costs associated with human-mediated question answering on forums. Conventional learning recognizes the resource-intensive nature of human responses. In contrast, the efficiency and time-saving potential of chatbots, particularly those rooted in the FAQ framework, offer an appealing solution. By employing chatbots, organizations can substantially offset the costs incurred by human intervention while simultaneously delivering prompt and tailored responses to user queries. This cost-effectiveness, coupled with the growing influence of AI across industries, underscores the current research enthusiasm in the realm of chatbot FAQs.

\begin{figure}[h!]
\centering
\includegraphics[width=0.8\textwidth]{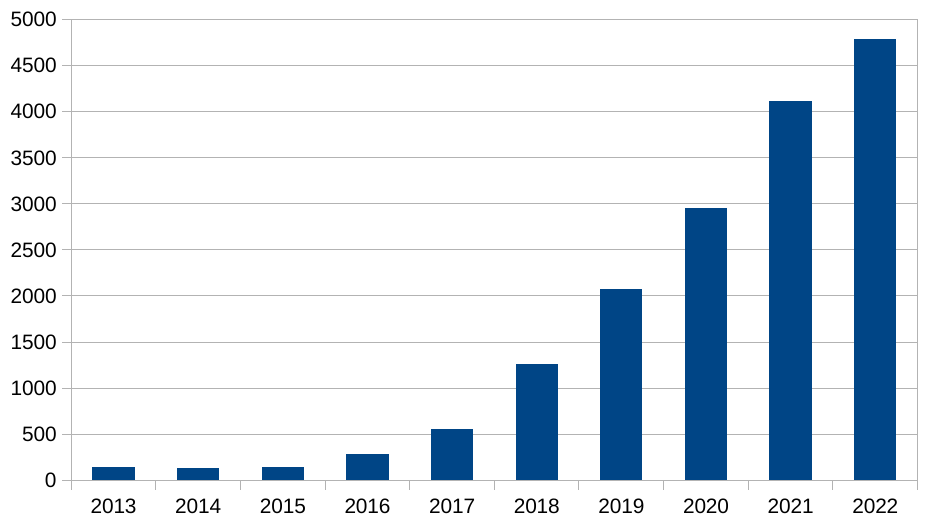}\\
\caption{\textcolor{black}{Progression of the number of works related to FAQ Chatbot over the last 10 years.}}  
\label{fig:chart10}
\end{figure}

 This study is motivated by the recent developments in the technology of chatbots, as ChatGPT became the most popular and useful chatbot after its release in 2022. There are some reviews available on other chatbots and ChatGPT; however, to the best of the author's search and efforts, a review exclusively exploring every component and main contribution in this field could not be found.

\begin{figure}[h!]
    \centering
    \includegraphics[width=0.8\textwidth]{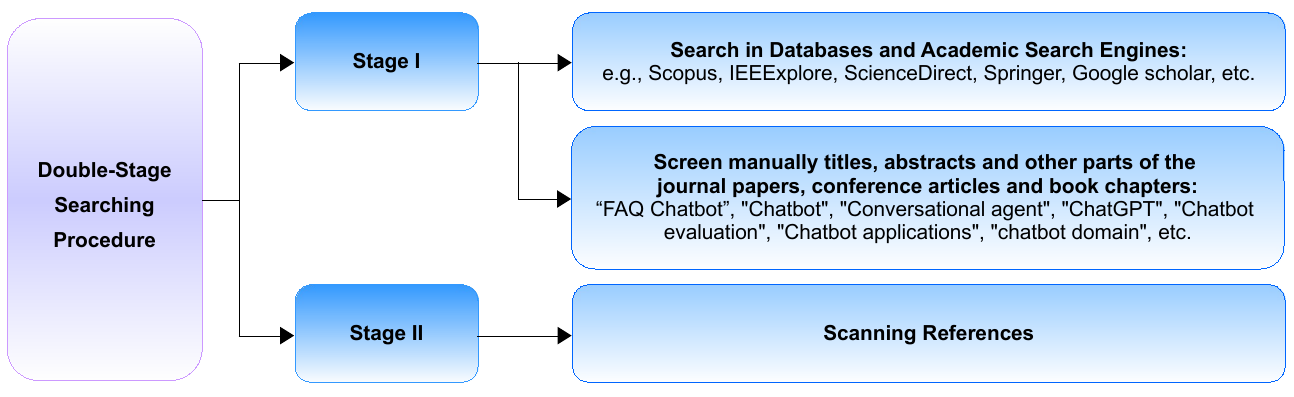}\\
    \caption{\textcolor{black}{Adopted search procedure for review.}}
    \label{fig:SearchProcess}
\end{figure}

\begin{figure}[h!]
    \centering
    \includegraphics[width=0.9\textwidth]{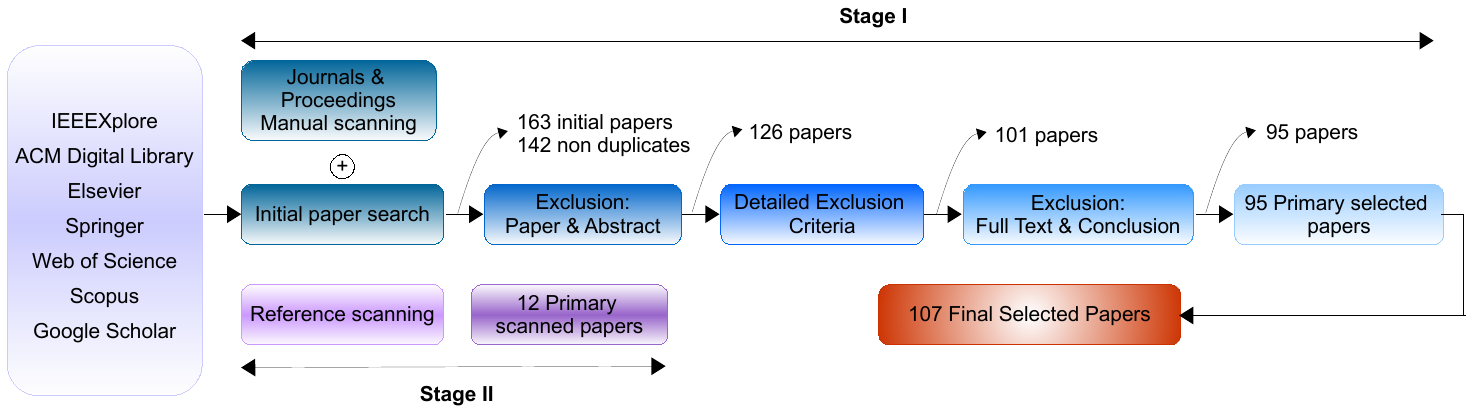} \\
    \caption{\textcolor{black}{Selection criteria of the research article for the presented review }}
    \label{fig:SelectionCriteria}
\end{figure}

\textcolor{black}{Typically, in response to the relevant queries, 163 articles were retrieved, with 142 of them being non-duplicates. These articles underwent further scrutiny to determine their relevance to the theme of the present study. Ultimately, 107 papers were selected for inclusion in this study.}

\section{Chatbots Background and Scope of Research}
\begin{figure}[h!]
    \centering
    \includegraphics[width=0.9\textwidth]{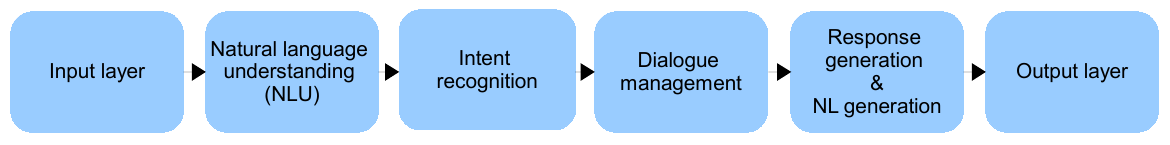}\\
    \caption{\textcolor{black}{NLP chatbot schema.}}
    \label{fig:NLP}
\end{figure}

"Can a machine think" - Alan Turing 1950, it started from this reflection, of machines being able to be autonomous. The idea of chatbots is to have a system that is capable of responding to questions automatically without human intervention. Chatbot implementation very often uses artificial intelligence methods. Due to significant advances in this field over the last few years, advanced chatbots are replacing humans in roles like customer support, online shopping assistants, and school administration. Ongoing tech progress promises wider adoption and improved interactions. These chatbots evolve with NLP and machine learning, excelling in FAQs, reducing human workload, and enhancing user experience. Making computers understand human language using NLP, allows pleasant dialogue between the machine and human. NLP is essentially used to create an Intelligent chatbot capable of responding to questions, without human intervention. Figure.\ref{fig:NLP} represents the overall architecture of an NLP-based chatbot. First, the input layer receives the user's input. Then, the natural language understanding component processes the input to extract the user's intent and extract important entities. The intent recognition component determines the user's intention based on their input, which is used to determine the appropriate response. The dialogue management component then determines the appropriate response and generates natural language output. Finally, the output layer provides the chatbot's response to the user. In another study \cite{chao2021emerging} authors survey the NLP chatbot approach, that allowed the implementation of intelligent chatbots. They study some intelligent text-mining techniques for key terminology extractions, the clustering method for identifying the subdomains, and the Latent Dirichlet Allocation for finding the key topics of the patent set. this research utilizes the Derwent Innovation database as the main source for global intelligent chatbot patent retrievals.


ALICE, which stands for Artificial Linguistic Internet Computer Entity, Early AIML chatbot, rule-based, by Dr. Richard Wallace in 1995. In \cite{abushawar2015alice}, Authors evaluate ALICE chatbot, which excels in diverse topics and has high flexibility. The study reveals that ALICE excels in answering general knowledge questions but is less proficient in specific technical domains. The research underscores the significance of maintaining a high-quality dataset and the necessity for regular updates and supervision to uphold accuracy and relevance. In tandem, there's an ongoing exploration of Frequently Asked Questions chatbots, with numerous studies delving into this domain. In \cite{mnasri2019recent} authors made the first chatbot, baptized ELIZA. This chatbot Acts as a psychotherapist and uses keyword-matching for patient interaction. Rules-based, context-dependent, challenging for new fields  \cite{caldarini2022literature}. After the advent of AI and advances in the field of computing at the beginning of the 21st century, several chatbots have been implemented \cite{caldarini2022literature} \cite{nithuna2020review}. The advent of AI has significantly transformed chatbots. Once constrained in understanding and responding, AI's inclusion with NLP, machine learning, and deep learning has elevated their intelligence. Chatbots now permeate education, functioning as school FAQs, aiding online courses, and streamlining administrative tasks. Furthermore, the positive impact and feedback of chatbots in the education sector have led to the consideration of implementing a chatbot to replace the existing FAQ system at ESTIN\footnote{\url{https://estin.dz/}} higher school FAQ as part of this state-of-the-art initiative.

\subsection{Taxonomy}
\begin{figure}[h]
    \centering
    \includegraphics[width=0.95\textwidth]{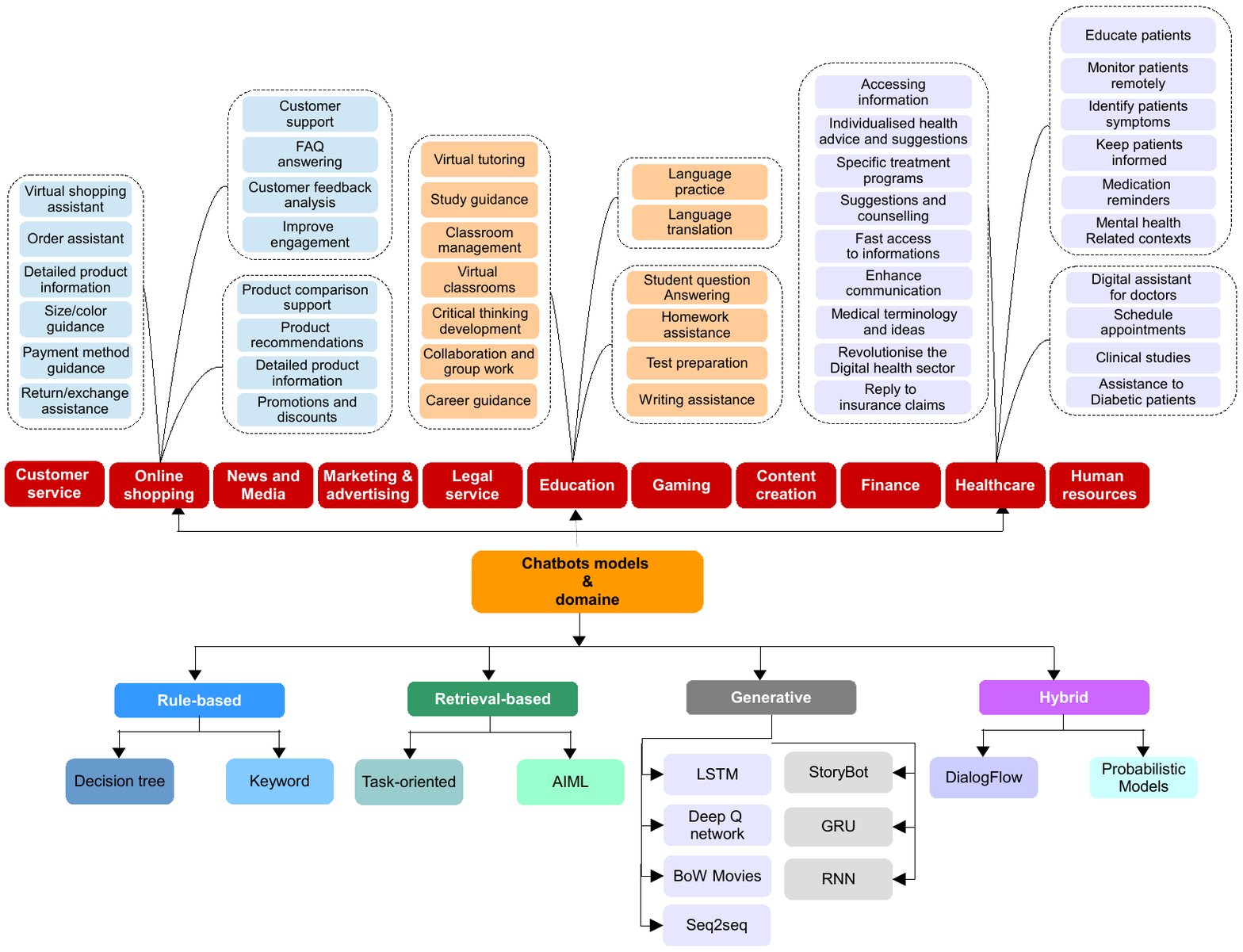}\\
    \caption{\textcolor{black}{Chatbot taxonomy.}}
    \label{fig:taxonomyG}
\end{figure}

To develop a systematic categorization of AI chatbots used for FAQs, the objective of this taxonomy should be first determined. Key considerations include understanding the underlying technology, the domain or industry of the application, the user experience offered, the source of the bot's knowledge, its platform of operation, and its monetization model. It is also vital to visualize this taxonomy effectively and ensure it is updated regularly to stay relevant in the fast-evolving AI landscape. Figure. \ref{fig:taxonomyG} portrays the taxonomy of existing AI chatbots used for FAQs.

\subsection{Chatbot evaluation metrics}
The rising prevalence of chatbots in recent years has seen increased design efforts by enterprises. Despite the surge, chatbot evaluation metrics remain inconsistent and often overlooked. While building a chatbot is commendable, its efficacy is determined by its ability to deliver accurate responses to users. \cite{adiwardana2020towards} introduced a human evaluation metric for multi-turn, open-domain chatbots, emphasizing the significance of diverse metrics. Meanwhile, \cite{liu2016not} highlighted the inadequacy of repurposing machine translation metrics for chatbot response systems, noting their weak correlation with human evaluations. The authors also spotlighted gaps in current metrics, advocating for improved chatbot evaluation metrics. In another study of \cite{shawar2007different} authors used three main chatbot performance metrics. Efficient evaluation of chatbots includes matching, response quality metrics, and user satisfaction, and adapting evaluation to the application and user needs is recommended. A user-friendly interface is as vital as a robust model, it boosts engagement and effectiveness. See Table \ref{tab1} for chatbots' evaluation metrics summary.


\begin{center}
\scriptsize
\begin{longtable}[!t]{
m{1.5cm}
m{2.9cm}
m{5cm}
m{5cm}}
\caption{Summary of Chatbots' evaluation metrics.}
\label{tab1}
\\ \hline
Reference & Metric & Description & Advantage/Limitation  \\ \hline
\endfirsthead
\multicolumn{4}{c}{{Table \thetable\ (Continue)}} \\
\hline
Reference & Metric & Description & Advantage/Limitation \\  \hline 
\endhead
\hline
\endfoot

 \cite{liu2016not} (2016) \newline \cite{mathur2020tangled} (2020) \newline \cite{caldarini2022literature} (2022) \newline  \cite{rosario2023grading} (2023) \newline \cite{khayrallah2023choose} (2023) & Bilingual Evaluation Understudy (BLEU) & 
     \textbullet~BLEU assigns a 0 to 1 value to translations. \newline \textbullet~A higher BLEU score reflects human-like translations
 . & 
     \textbullet~ Prioritizes overlap, and neglects response quality, coherence, and appropriateness. \newline \textbullet~ Lack of human judgment and subjective assessment of translation quality. \newline \textbullet~ Quantitative measure for translation comparison and system evaluation. \newline \textbullet~ Assesses similarity between translations and references.
 \\

 \cite{liu2016not} (2016) \newline \cite{rosario2023grading} (2023) \newline \cite{khayrallah2023choose} (2023) & Metric for Evaluation for Translation with Explicit Ordering (METEOR) & \textbullet~ Originally for machine translation, adaptable to chatbots. \newline \textbullet~ METEOR assesses translations lexically, syntactically, and linguistically. 
 \newline \textbullet~ Requires human-generated reference responses for chatbot evaluation. 
 \newline \textbullet~ Scores similarity through n-gram alignment and linguistic features. & 
 \newline \textbullet~ Requires a reference set of human-generated responses for comparison, limiting its applicability without suitable references. 
 \newline \textbullet~ Subjectivity in determining linguistic weights and penalties.  
 \newline \textbullet~ Enhance correlation with human judgment compared to BLEU. \
 newline \newline \textbullet~ Improves translation evaluation by penalizing incorrect word order.\newline\\

\cite{liu2016not} (2016) \newline \cite{khayrallah2023choose} (2023) \newline \cite{rosario2023grading} (2023) & Recall Oriented Understudy for Gisting Evaluation (ROUGE) & 
\newline \textbullet~ Measures summary/response quality vs. reference texts. 
\newline \textbullet~ Focuses on recall, assessing n-gram overlap. 
\newline \textbullet~ ROUGE-N for various n-gram lengths, ROUGE-L for common subsequence. 
\newline \textbullet~ ROUGE-W weighs based on match length, ROUGE-S for skip-bigrams. 
\newline \textbullet~ Score ranges 0 to 1, 1 signifies perfect match to reference. & 
\newline \textbullet~ Primarily designed for summarization evaluation, not specifically for chatbot assessment. 
\newline \textbullet~ Subjective decisions may be required in setting appropriate n-gram lengths or weighing factors. 
\newline \textbullet~ Supplies a quantitative measure of overlap between generated summaries/responses and reference texts. 
\newline \textbullet~ Proposes multiple variants (ROUGE-N, ROUGE-L, ROUGE-W, ROUGE-S) to capture different aspects of recall.\newline \\
                   
\cite{adiwardana2020towards} (2020) & Static evaluation & 
\newline \textbullet~ Static evaluation assesses chatbot performance without user interaction. 
\newline \textbullet~ Involves examining design, functionality, content, and language processing. \newline \textbullet~ Utilizes predefined criteria for evaluation. & 
\newline \textbullet~ Limited assessment without user interactions.
\newline \textbullet~ Pre-deployment issue identification aid. \newline\\

                   & Measuring human likeness & 
                   \newline \textbullet~ Evaluating chatbot human-likeness involves comparing its responses to human ones. & 
                   \newline \textbullet~ Subjective nature 
                   \newline \textbullet~ Improve engagement. 
                   \newline \textbullet~ Building trust. \newline  \\

                  & Perplexity for automatic evaluation & 
                  \newline \textbullet~ Metric in NLP and chatbot assessment. 
                  \newline \textbullet~ Gauges language model's word sequence prediction.  
                  \newline \textbullet~ Assesses response fluency and coherence. 
                  \newline \textbullet~ Measures word probability assignment based on context. & \newline \textbullet~ Restricted to language modeling only. 
                  \newline \textbullet~ Limited sensitivity to semantic meaning. 
                  \newline \textbullet~ Evaluation of linguistic proficiency. \newline \\  

                   & Interactive evaluation & 
                   \newline \textbullet~ Interactive chatbot evaluation involves user interaction. 
                   \newline \textbullet~ Real-time conversations gather feedback on answer quality. \newline \textbullet~ Assesses user satisfaction and overall experience. & 
                   \newline \textbullet~ Time-consuming and resource-intensive. 
                   \newline \textbullet~ Influenced by user expertise and familiarity. 
                   \newline \textbullet~ Enhanced insight into chatbot capabilities.
                   \newline \textbullet~ Real-time performance feedback. \newline \\     

\cite{peras2018chatbot} (2018) & Usability  & 
\newline \textbullet~ Chatbot usability assesses efficiency and effectiveness. 
\newline \textbullet~ focuses on ease of use and task completion time. 
\newline \textbullet~ Metrics are often qualitative, involving user characteristics. 
\newline \textbullet~ Leverages user experience and perspectives. & 
\newline \textbullet~ Usability evaluation is subjective due to varying user perceptions and interpretations. 
\newline \textbullet~ Enhance user experience. \\

                        & Satisfaction  & 
                        \newline \textbullet~ Users' pleasure based on expectations. 
                        \newline \textbullet~ Tied to chatbot's performance. 
                        \newline \textbullet~ Some satisfaction metrics are quantitative. 
                        \newline \textbullet~ Studies examine satisfaction in relation to chat duration, and frequency. & 
                        \newline \textbullet~ Lack of standardized metrics. 
                        \newline \textbullet~ Difficulty in quantifying satisfaction. 
                        \newline \textbullet~ User-centered judgment. \newline\\    

                        & Accuracy  & 
                        \newline \textbullet~ Chatbot accuracy evaluation matches answers to references. 
                        \newline \textbullet~ Determines correctness and precision. 
                        \newline \textbullet~ Assesses information delivery and task solving. & 
                        \newline \textbullet~ Reliance on reference answers or expert judgments. \newline \textbullet~ Objective assessment. \newline \\    
                        
                        & Accessibility  & 
                        \newline \textbullet~ Chatbot accessibility evaluation considers usability for diverse users. 
                        \newline \textbullet~ Ensures usability for individuals with disabilities.
                        \newline \textbullet~ Includes visual, hearing, and cognitive impairments.
                        \newline \textbullet~ Examines design, interface, and functionality. 
                        \newline \textbullet~ Aims to identify and address accessibility obstacles. & 
                        \newline \textbullet~ Technical Limitations. 
                        \newline \textbullet~ Promotes equal access for all users. \\

\end{longtable}
\end{center}


\section{Chatbot approaches overview: Taxonomy of existing methods} 
After the advent of the well-known ChatGPT, chatbots have become necessary for any business. Since they are used in different applications, including customer service, virtual assistant, and education. With different types of chatbots and models available, it is a big challenge to select the adequate one for the application. To assist the users, it has been decided to present a taxonomy of this model in this paper. This taxonomy categorizes chatbots into various types based on the models used. These categories include rule-based chatbots, retrieval-based chatbots, generative chatbots, intent-based chatbots, and hybrid chatbots. Each category encompasses different models and techniques, each with its unique advantages and disadvantages. Having an understanding of the various types of chatbot models allows businesses and developers to choose the most fitting model for their specific needs. As a result, chatbot solutions can become more effective and efficient. Many well-known chat dialogue systems are built using one or multiple approaches. This section introduces various chatbot designs and methodologies, ranging from rule-based chatbots, retrieval-based chatbots, and generative chatbots to hybrid chatbots. A fundamental contrast between rule-based chatbots and retrieval-based chatbots is the approach they use to produce responses to user inputs. Rule-based chatbots generate responses according to predetermined rules, whereas retrieval-based chatbots find the most appropriate response by matching user inputs to existing responses in a database.


\subsection{Rule-based chatbots}
\begin{figure}[!h]
\centering
\includegraphics[width=0.9\textwidth]{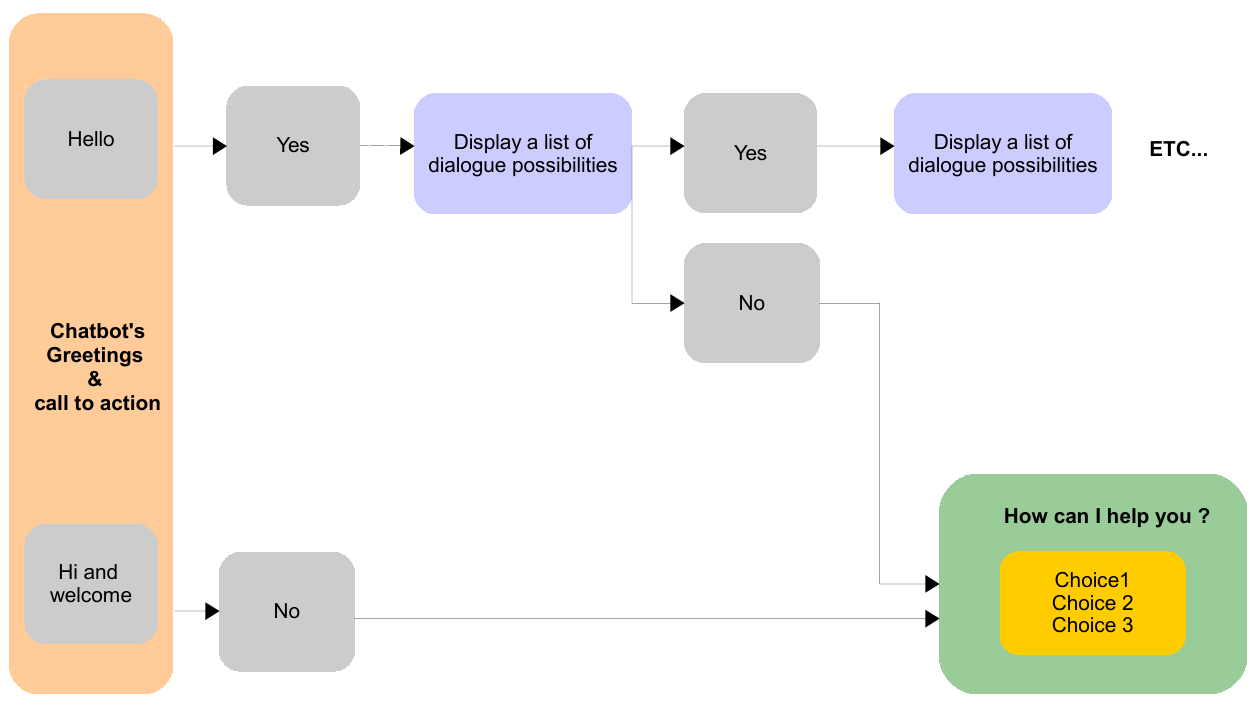}\\
\caption{\textcolor{black}{Rule-based chatbot model.}}
\label{fig:decisionTree}
\end{figure}

Rule-based chatbots follow predefined rules using decision trees or flowcharts for limited, pattern-based responses, lacking flexibility compared to other chatbot types. Figure. \ref{fig:decisionTree} shows the general scheme of a tree decision chatbot. As observed, the chatbot typically initiates the conversation with greetings and calls to action, aimed at encouraging the user to express their request. This type of chatbot uses Yes or No answers and/or displays choices to the user, so as to frame the conversation and predict the course of the conversation. This section covers two types of rule-based chatbots: decision tree chatbots and keyword chatbots. Decision tree chatbots follow steps like welcome messages, user option selection, information retrieval, and repetition. Keyword chatbots prompt user input, analyze for specific keywords, and respond accordingly. While rule-based systems lack complexity, they ensure consistent and predictable responses, fostering user trust for specific use cases.


\subsubsection{Decision tree chatbots}
Decision tree chatbots are the most basic type of chatbot. These chatbots are rather simple to implement because they are based on pre-established scenarios. Therefore, the main issue of these chatbots is that they can very quickly fall into the generic case where only one answer is returned to the user, which will lead him to abort the discussion with the chatbot. In \cite{mujeeb2017aquabot}, a chatbot named "Aquabot" was designed to assess the severity of disease from user text queries. Utilizing NLP for understanding and Decision Trees for patient classification, the chatbot has three primary modules: replying to users, processing inputs through a decision tree, and the decision tree itself. With an 88\% accuracy rate compared to human psychologists, it offers efficiency and resource-saving benefits, making it suitable for aiding psychologists. On the other hand, \cite{saha2022healthcare} introduced a healthcare chatbot that supersedes traditional FAQs, offering real-time answers to patients' questions. This model is trained on an extensive database of symptoms and diseases, using a decision tree approach to guide interactions and assess patients' conditions. This interactive process ensures more precise diagnoses. Overall, this system enhances virtual assistant FAQs by providing a user-centric and highly accurate diagnostic experience leveraging a robust database and decision tree methodology. During the COVID-19 crisis, many entities used chatbots to replace human assistants. In \cite{mellado2020learning}, a study explored accounting students' experience with a rules-based chatbot, using decision trees, for tax control instruction during remote learning. The findings indicate that this approach was more effective than other remote learning methods, highlighting its potential in areas traditionally resistant to innovative teaching methods.

\subsubsection{Keyword chatbots}
Keyword chatbots are a kind of rule-based chatbots. The system looks for a keyword in the user's input, in order to provide the appropriate response. This type of chatbot uses specific keywords or phrases to identify the user's intent and respond appropriately. In \cite{tsai2019ask}, authors propose a keyword-based chatbot system, named Ask Diana, for water-related disaster management. Ask Diana is comprised of three primary modules: the water-related disaster database, the user intent comprehension mechanism, and the user interface. The database collects both dynamic and static data concerning water disasters. As a chatbot system, Ask Diana's user interface was designed with chatbot characteristics in mind, and the data presentation method was developed to effectively communicate information on a mobile device. The user intent comprehension mechanism is responsible for identifying the user's desired information or data. This mechanism utilizes a fuzzy search algorithm to analyze the user's input text and a keyword table to match the inputted keywords with the information stored in the database.
Table \ref{tab2} summarizes the rule-based chatbots for FAQs.

\begin{center}
\scriptsize
\begin{longtable}[!t]{
m{1.1cm}
m{1.5cm}
m{2cm}
m{2cm}
m{3.6cm}
m{3.9cm}}
\caption{Summary of rule-based chatbots for FAQs.}
\label{tab2}
\\ \hline
 Work & Model & Dataset & Application & Performances & Advantage/Limitation   \\ \hline
\endfirsthead
\multicolumn{6}{c}{{Table \thetable\ (Continue)}} \\
\hline
Work & Model & Dataset & Application & Performances & Advantage/Limitation  \\ \hline 
\endhead
\hline
\endfoot

\cite{mujeeb2017aquabot} (2017) & Decision tree &  N/A  & Medical field & precision = 88\% &   \textbullet~Lacks detailed evaluation and validation. \newline \textbullet~Accessibility and convenience.  \newline\\

\cite{saha2022healthcare} (2022) & Decision tree &  Real users  & Medical field & The chatbot provides precise diagnoses for users' health issues based on their queries. &   \textbullet~Lack of evaluation and validation details. \newline \textbullet~Easy to implement.  \newline\\

\cite{mellado2020learning} (2020) & Decision tree &  N/A  & Education &  Improvement \newline Male = 27.8 \% \newline Female = 62.5 \%  &  Limited scope \newline \textbullet~Lack of evaluation and validation \newline \textbullet~Efficient information retrieval.  \newline\\

\cite{tsai2019ask} (2017) & Keyword &  Real users \& simulated disasters.  & Water-related disaster management &  Efficiently access disaster data, develop strategies. &  \newline \textbullet~ Limited scope \newline \textbullet~ Efficient information retrieval.  \newline\\

\end{longtable}
\end{center}

\subsection{Retrieval-based chatbots}
Retrieval-based chatbots are a category of chatbot models that utilize a predefined set of responses stored in a database based on known questions and answers. These chatbots operate by matching user input with pre-existing responses in the database, rather than generating novel responses from scratch like generative chatbots. In order to create this kind of chatbot, the first step involves gathering a large dataset of potential questions and corresponding answers. This dataset will be then used to match the question user's input and return its response. Retrieval-based chatbots can be effective in situations where the questions and answers are straightforward and consistent. However, they may struggle with more complicated or nuanced queries since they are limited to the pre-defined responses in their database. contrariwise retrieval-based chatbots have a relatively simple design, which makes them easier to develop and maintain compared to more advanced chatbot models. They can also provide prompt and dependable responses to common questions, making them useful for applications such as customer support or information provision.

\subsubsection{Artificial Intelligence Markup Language} \label{ctp}
\begin{figure}[h!]
    \centering
    \includegraphics[width=0.9\textwidth]{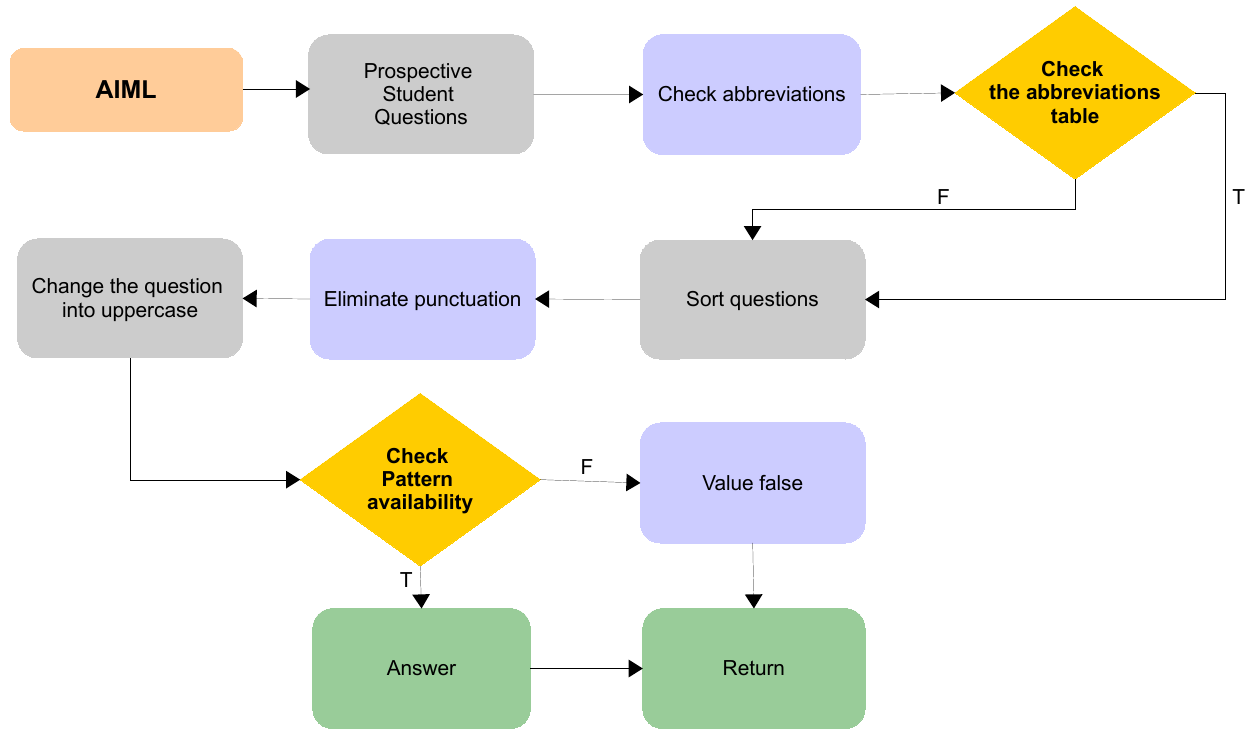}\\
    \caption{\textcolor{black}{AIML flowchart \cite{wijaya2020chatbot}.}}
    \label{fig:aimlchart}
\end{figure}

The Artificial Intelligence Markup Language (AIML) was created from 1995 to 2000, and it is based on the concepts of Pattern Recognition or Pattern Matching technique \cite{adamopoulou2020overview}. It is used in natural language modeling for conversation between humans and chatbots using the stimulus-response approach. It is an XML-based markup language and it is tag-based. AIML is based on basic units of dialogue called categories (tag <category>) which are formed by user input patterns (tag <pattern>) and chatbot responses (tag <template>) \cite{adamopoulou2020overview}. This approach gained popularity after being used in the successful dialogue system A.L.I.C.E. \cite{shawar2002comparison} that won the Loebner Prize three times \cite{yamaguchi2018chatbot}. Since then, AIML has been adopted in many works. Authors in \cite{wijaya2020chatbot} analyze the design and implementation of a chatbot for supplying information related to new student registration. The chatbot was designed using AIML (Artificial Intelligence Markup Language) and machine learning techniques. The AIML flowchart for this chatbot as implemented in this paper is shown in the figure.\ref{fig:aimlchart}. In \cite{ranoliya2017chatbot}, an AIML-based chatbot was created for Manipal University to assist students in obtaining information. The chatbot first uses AIML templates to find answers by seeking similarities with user queries. If AIML can't find an answer, Latent Semantic Analysis (LSA) steps in to determine word similarities and provide a response. The chatbot, however, requires regular updates to maintain its relevance. Similarly, \cite{khin2020university} designed an AIML chatbot for Yagon University. Hosted on Pandorabots servers, the chatbot underwent testing and displayed accuracy in most responses, with a few errors attributable to user input or chatbot knowledge. Incorporating enhanced error management is suggested to improve efficiency, as noted in \cite{buhrke2021making}. Moving on, \cite{thomas2016business} presented an e-business chatbot for customer service using AIML and LSA. The paper delineated the integration process and highlighted the benefits of combining AIML and LSA, emphasizing improved language understanding. However, it didn't offer a thorough evaluation or compare the model with other chatbot systems.

In \cite{mikic2018using}, AIML was employed to create a chatbot aiding students in computer Architecture courses at the University of Vigo (Spain). This chatbot effectively fielded frequent questions, conserving teachers' time. The paper outlines the chatbot's design, development, and the creation of a knowledge base to ensure efficient student interactions. The chatbot significantly improved students' learning experiences, offering clarity and additional support beyond class hours.

\subsubsection{Task-Oriented Chatbots}
These chatbots are designed to perform specific tasks for the user, such as booking a hotel room or ordering food. They work by matching user input to pre-existing workflows or scripts. In \cite{li2020conversation}, authors analyze the interaction between users and a banking chatbot that is designed to assist with specific tasks. The study uses conversation analysis to examine instances where the chatbot is unable to complete the user's request, and the user employs coping strategies to deal with the issue. This study aims to find where the chatbot is unable to complete the user's request due to system limitations or user errors. The authors identify several coping strategies that users employ in these situations, such as repetition, rephrasing, and simplifying their requests. They also analyze instances where the chatbot uses a non-progress strategy, including changing the subject or providing irrelevant information, to avoid admitting that it can not fulfill the user's request. The paper provides valuable insights into how users interact with task-oriented chatbots and the coping strategies they employ when these systems are unable to complete their requests. The study highlights the need for chatbot designers to consider the limitations of their systems and the strategies users employ to deal with them.
Table \ref{tab3} presents a summary of retrieval-based chatbots for FAQs.

\begin{center}
\scriptsize
\begin{longtable}[!t]{
m{1.1cm}
m{1.2cm}
m{1.6cm}
m{1.5cm}
m{3cm}
m{5.5cm}}
\caption{Summary of retrieval-based chatbots for FAQs.}

\label{tab3}
\\ \hline
 Work & Model & Dataset & Application & Performance & Advantage/Limitation   \\ \hline
\endfirsthead

\multicolumn{6}{c}{{Table \thetable\ (Continue)}} \\
\hline
Work & Model & Dataset & Application & Performance & Advantage/Limitation  \\ \hline 
\endhead

\hline
\endfoot

\cite{thomas2016business} (2016) & AIML &  Real users & Customer service &  Precision = 0.97 &
\newline \textbullet~ Lacks a comprehensive evaluation of the chatbot's performance and user satisfaction.  \newline\\

\cite{ranoliya2017chatbot} (2017) & AIML &  Real users  & Education & N/A &
\newline \textbullet~ Specific domain. 
\newline \textbullet~ Does not explore scalability or adaptability. 
\newline \textbullet~ Reduce the workload of university staff. \newline\\

\cite{mikic2018using} (2018) & AIML &  Real users & Education &  System reduces teacher Q\&A, supports student doubts.  &
\newline \textbullet~ Lacks a comprehensive evaluation of the chatbot's performance and user satisfaction. 
\newline \textbullet~ It saves teachers' time and helps students to get speedy answers. \newline\\

\cite{khin2020university} (2020) & AIML &  Real users & Education &  N/A & 
\newline \textbullet~ May struggle to handle more complex or nuanced questions. 
\newline \textbullet~ Lack of evaluation details. \newline \\

\cite{wijaya2020chatbot} (2020) & AIML & 100 Q\&A & Education & Precision = 77\% \newline Recall = 87,5\% \newline Accuracy = 90\% &
\newline \textbullet~ Does not explore scalability.
\newline \textbullet~ It saves administrators' time and helps students to get speedy answers. \newline\\

\cite{li2020conversation} (2020) & Task-oriented & 24,074 exchanges  & Banking & N/A &
\newline \textbullet~ Repeat the same request.
\newline \textbullet~ Rephrasing the question. 
\newline \textbullet~ Switching different topics.
\newline \textbullet~ Specific context, detailed analysis. 
\newline \textbullet~ More understanding of user-bot interactions. \newline \\

\end{longtable}


\end{center}

\subsection{Generative models}
Generative chatbots offer a unique advantage of providing more interactive and customized conversations with users as they are not confined to pre-existing responses and can adapt to new situations and contexts. Nonetheless, building and maintaining them requires a significant amount of training data and computational resources. Furthermore, since the model creates responses from scratch, there is a possibility of generating irrelevant or unsuitable responses. Hence, meticulous testing and monitoring are critical to guarantee the quality and safety of generative chatbots. Chatbots using the generative model aim to generate autonomous word-by-word responses, in order to conduct a conversation in a more independent way. For more efficiency, this model can be combined with Neural Network Models \cite{serban2016building}. To train a generative chatbot, a large dataset of questions and answers is typically required, or at the very least, the establishments expressing the need for a chatbot must possess a small domain-specific dataset (such as an FAQ) about their products, services, or used technologies \cite{kapovciute2020domain}.

\subsubsection{Recurrent Neural Network }

RNN chatbots use Recurrent Neural Networks (RNN) to generate context-aware responses from sequential data, enhancing naturalness, and engagement in conversations. They're vital for handling lengthy sequences, with data continuously fed into the network cells. Long Short-Term Memory neural networks (LSTM) are RNNs that allow conversational agents to distinguish between long and short-term memory. Long-term memory refers to the model's general knowledge data for conversation prediction and generation, whereas short-term memory refers to knowledge data that is only valid and relevant for a specific time window defined by recent user-agent interaction \cite{motger2022software}. Authors in \cite{serban2017hierarchical} introduce a hierarchical latent variable RNN for complex, multi-level generative processes in sequential data like Twitter conversations. The model attaches high-dimensional latent variables to dialogue utterances, generating responses by sampling these variables and producing words accordingly. The chatbot model is evaluated and compared to competing models via a human evaluation study on Amazon Mechanical Turk. The experiments show that this chatbot model enhances upon recently proposed models and that the latent variables simplify the generation of meaningful, long, and diverse responses and maintain dialogue context. In another hand, researchers in \cite{xu2017new} created a chatbot for customer service on social media. Long short-term memory (LSTM), a particular type of RNN, was used to generate responses for customer service requests on social media. The system processes the requests from customers to vector representations, feeds them to LSTM, and then reports the response. The system was trained on nearly 1M Twitter conversations between users and agents from 60+ brands. The results showed that this chatbot is as good as humans, in terms of showing empathy to assist users, and it exceeds the information retrieval system founded, on both human judgments and an automatic evaluation metric. In another work, and in order to remedy the absence of support in colleges during the weekends, and especially at the time of admission, authors in \cite{huddar2020dexter} create a chatbot that is able to answer multiple persons at the same time, so people don't have to visit the college to get their query solved and it will be available 24/7. This chatbot is developed using RASA stack, and it is RNN-based because it needs to be able to frame its answer if the answer for a particular question is not available in the database.



\subsubsection{GRU Question Generator}
GRU, which stands for Gated Recurrent Unit, is a type of neural network architecture that is commonly used for generating questions within question-answering systems (FAQ) or chatbots. This architecture is similar to the better-known LSTM (Long Short-Term Memory) architecture and is specifically designed to handle sequential data like text. The GRU model is utilized in question generation tasks to predict a relevant question based on the given input text or context. In the context of \cite{kamphaug2018towards} the focus is on crafting a data-driven conversational chatbot utilizing the GRU model.. The preoccupation of the research is to show the challenge of allowing chatbots to effectively participate in open-domain conversations and generate responses that are meaningful and relevant. The study emphasizes the need for context-aware chatbot models, proposing the GRU architecture known for handling sequential data. It trains by encoding input and generating output with GRU networks, assessing performance using metrics like relevance and coherence. This approach marks a significant advance in conversational AI, paving the way for future research and improvements. Authors in \cite{liu2019learning} introduce a model called Clue Guided Copy Network for Question Generation (CGC-QG). This model is based on an encoder-decoder framework with attention and coping mechanisms, including a clue word predictor, passage encoder, and question decoder. The predictor helps identify possible clue words in the input passage that can be copied into the output question. For the passage encoder component, the researchers used a bidirectional Gated Recurrent Unit (BiGRU), and this step combines multiple elements, including the predicted clue word distribution and various feature embeddings of the input words. During the decoding stage, the authors also employed a separate GRU (Gated Recurrent Unit) with a coping mechanism. This GRU generates the question words one by one, using the encoded representation of the input passage and the precedent-decoded words as references. This model intelligently selects and generates words from the input passage, enhancing question coherence and relevance, essential for human-like conversational chatbots. In another work \cite{wang2019answer} authors recognize that the answer provided in a conversation shows strong semantic coherence with its corresponding question. Considering this insight, they propose two methods to enhance the semantic coherence between the post and to generate questions, driven by its answer. For the first method, a reinforcement learning framework is employed, to maintain the coherence between the question and its related answer. They also present a novel model called GRU-MatchPyramid, which improves the MatchPyramid model by incorporating a bi-directional GRU. By doing so, the researchers aim to capture higher-level semantic information about words. They also show in this article that these findings contribute to advancing the field of question generation and provide valuable insights for developing chatbots capable of generating contextually appropriate and engaging questions.



\subsubsection{Seq2seq Model}

Seq2seq model chatbots use deep learning to produce human-like responses by employing encoder and decoder neural networks. The encoder processes user input into a fixed-length vector, and the decoder generates responses word by word. They've gained popularity for their natural responses compared to rule-based chatbots and can improve with large conversational datasets. However, they demand substantial high-quality data and can face challenges maintaining coherence and relevance in conversations due to word-by-word generation. Authors in \cite{adiwardana2020towards} present \textit{Meena}, a chatbot based on the generative model, and trained end-to-end on 40B words filtered and come from social media conversations. This limitation of the end-to-end approach was pushed out of boundaries, in order to show that a large-scale low-perplexity model can have quality language results. Seq2seq model and the Evolved Transformer \cite{egonmwan2019transformer} were also used as the principal architecture. The training phase goes throw multi-turn dialogues, where the input sequence is all turns of the context (up to 7) and the output sequence is the response. To evaluate Meena and demonstrate its contribution over the remains of chatbots, the authors also proposed new performance metrics, which are Sensibleness and Specificity Average (SSA). These metrics include two basic qualities of a human-like chatbot: giving meaning and being specific in the response the chatbot gives. Overall, the article highlights the possibility of Seq2Seq models for developing more human-like chatbots that can be employed in open-domain conversations. However, it also emphasizes the challenges that remain in creating chatbots that can truly replicate human conversation. In \cite{boussakssou2022chatbot} authors introduce midoBot an Arabic chatbot that uses the seq2seq model to allow conversational interactions with users through text. This approach implicates the establishment and testing of the model using the Tensorflow 2 deep learning framework, particularly employing diverse seq2seq model architectures. The experimental results demonstrate the effectiveness of the approach, as the chatbot consistently generates meaningful responses for a wide range of user questions. Across the board, this article contributes to the field of NLP and chatbot development by concentrating on the typical challenges and opportunities related to Arabic language chatbots. It underlines the prospect of Seq2Seq models in facilitating conversational experiences in Arabic and supplies insights for future research and improvements in this domain. Chatbots are offering valuable information and support in the educational domain, to students, teachers, and education staff. Authors in \cite{khin2020question} experiment with the application of a Seq2Seq model with an Attention Mechanism based on an RNN encoder-decoder model to boost communication in a neural network chatbot. Compared to rule-based chatbots, RNN chatbots are more popular and scalable. The chatbot in this research was designed for the university education sector, providing answers to frequently asked questions and offering relevant information. It is worth noting that this chatbot is the first of its kind in the Myanmar language, utilizing a neural network model and achieving a BLEU score of 0.41.

\begin{figure}[!h]
\centering
\includegraphics[width=1\textwidth]{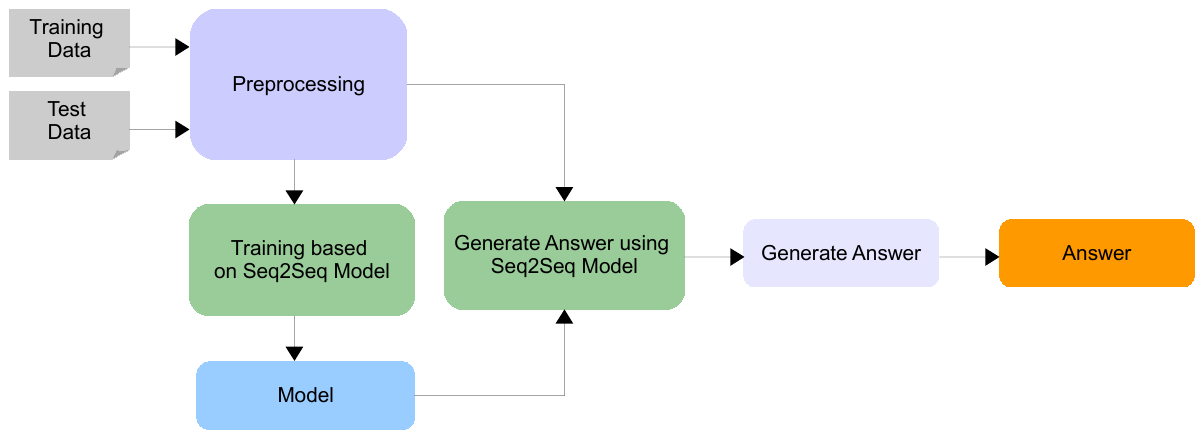}\\
\caption{\textcolor{black}{Seq2seq chatbot implemented in \cite{khin2020question}.}}
\label{fig:seq2seq}
\end{figure}

\subsubsection{Deep Q Network}
Deep Q Networks (DQNs) are a variety of machine learning algorithms that aim to reinforce learning. They have attained popularity due to their capacity to solve complex problems in areas such as robotics, and control systems. The basic idea behind DQNs is to utilize a deep neural network to calculate the value of actions in a given state, permitting the algorithm to learn how to make optimal decisions through trial and error. DQNs are able to adjust to changing environments and make decisions in real-time. However, this system is not well-suited for all types of problems and requires careful tuning and design to attain optimal results. Authors in \cite{ricciardelli2019self} present a study where RL is used to train a FAQ chatbot. The policy learning was addressed by the implementation of a Deep Q-Network (DQN) agent with epsilon-greedy exploration. This technique is particularly made to manage the challenge of providing fallback answers for out-of-scope questions in chatbot systems. By combining epsilon-greedy exploration, the DQN agent can intelligently balance between using the learned policy and exploring alternative actions, guaranteeing that the chatbot can handle questions that fall outside its field of expertise. DQN helped to enhance the general performance and robustness of the chatbot, qualifying it to provide significant responses.

\subsubsection{Storybot Model}
The StoryBot model is a famous method in the domain of conversational AI for creating chatbots that can engage in interactive and dynamic conversations with users. This bot uses a dialogue policy learned from a collection of dialogue samples or stories, to cater to the users' curiosity in storytelling, and generates brief fantasy stories upon user request. Joseph Sassoon, in the book entitled Storytelling and Artificial Intelligence: When Stories are Narrated by Robots \cite{sassoon2019storytelling} exposes the intersection of storytelling and artificial intelligence, by investigating how AI algorithms and techniques can be applied to storytelling. Exploring the effect on various fields such as advertising and education, it examines the role of AI in generating narratives, examining patterns, and understanding storytelling structures. Sassoon evokes storybot and discusses the implications of AI-generated narratives on creativity, human expression, and the future of storytelling. The book also studies ethical concerns and challenges associated with the use of AI in storytelling. StoryBot model was also used in \cite{bailey2021digital} to enhance language learning (L2)  that emphasizes narrative-based interactions. The authors of this research also aim to explore students' perceptions of these storybot exchanges and handle any shortcomings identified in earlier studies with regard to perception rates. The study shows that The integration of storybots in the L2 classroom produced varied participation rates, where students were engaging more in reading instead of writing activities. This indicates a strong priority on reading comprehension skills in storybot interactions. Survey results indicated that students perceived storybots as effective tools for achieving their L2 objectives, considering them appropriate to their language learning journey and user-friendly in terms of navigation. The authors also mentioned that, since the exchanges, in this study, were restricted to text messaging, no observations or investigation could be made concerning the effect on speech or pronunciation skills. Besides, the fact that it was completed within a specific university class in South Korea, can limit the generalizability of the findings to other regions or younger learners. In another work \cite{serban2017deep}, authors evoke the integration of a storyBot into their system, due to high requests from users who were interested in hearing stories from the socialbot proposed in this work. The Storybot detects if a user requests a story by analyzing the existence of specific words associated with requests and story types in the user's input. When a story is requested, the model generates a response that includes the title and author of the story, followed by the story body. Authors in \cite{xu2022elinor} introduce a conversational AI component into Elinor, a popular children’s science animation show. The purpose of this study is to make Elinor interactive, and thus enhance children's engagement and learning in science education, based on the dialogue learning experience. Through a sequence of design iterations and user studies, the authors evaluate the efficacy of integrating conversational AI into Elinor, and examine also how children interact with virtual characters, the impact of the conversational AI component on engagement and learning outcomes, and the perceptions of both children and teachers regarding the integration of AI. The results of the study indicate that the inclusion of conversational AI in Elinor positively influences children's engagement, motivation, and learning outcomes in narrative science programming. Children expressed excitement and a sense of connection when interacting with the virtual characters, and teachers realized the potential of conversational AI to enhance learning experiences.

\subsubsection{BoW Movies}
BoW Movies refers to the Bag-of-Words (BoW), which is a model that can also be used in the context of a Frequently Asked Questions (FAQ) chatbot that concentrates on movie-related inquiries. The BoW model can be utilized to process and apprehend user questions and provide pertinent responses based on a collection of movie-related frequently asked questions. This model is employed to illustrate the questions in the FAQ database by considering the occurrence and frequency of words. When a user asks a question to the chatbot, this model is used to compare the input question with the most suitable FAQ entry based on the similarity of word occurrences. Across the board, integrating the BoW model into a FAQ chatbot for movies enriches the chatbot's capacity to understand and reply to user questions effectively, improving the user experience and providing pertinent information about movies. Authors in \cite{nugraha2022chatbot} presented a movie recommender system that uses a chatbot and POS (Part-of-Speech) tagging. The chatbot in this study engages with users, comprehends their preferences, and suggests movies accordingly. The POS tagging approach is employed to analyze the user's text input and extract pertinent information for movie recommendation. The study underlines the usefulness of the chatbot-based approach in providing personalized movie recommendations and discusses the implementation details and evaluation results. The authors provide in this research personalized movie recommendations, whereas this system may be influenced by limited movie data. In \cite{bao2020plato} authors propose a training process for creating a high-quality open-domain chatbot called PLATO-2. This process implicates two steps. First, a coarse-grained generation model is trained to generate answers using a simplified framework. Second, a fine-grained generative model with latent variables and an evaluation model are trained to produce diverse responses and select the best one. The researchers also combine BoW with Latent variables in order to facilitate the training. The proposed model was trained in both Chinese and English data, and its efficacy was demonstrated through complete evaluations. 

\subsubsection{LSTM Classifier}
LSTM (Long Short-Term Memory) classifier is a class of Recurrent Neural Network (RNN) models, that is generally employed for sequence classification assignments. LSTM is particularly developed to manage long-term dependencies in sequential data. In the context of a FAQ chatbot, an LSTM classifier can be conditioned on a dataset of frequently asked questions (FAQs) and their related answers. The LSTM classifier learns to identify patterns and relationships in the input questions, allowing it to categorize new user queries into appropriate categories or topics. The classifier uses the LSTM's ability to capture long-term dependencies, helping it to understand the context and semantics of the questions and provide accurate responses. Thereby including an LSTM classifier in a FAQ chatbot, it can effectively handle a wide range of user queries and provide suitable responses based on the learned patterns. This method improves the accuracy and efficiency of the chatbot, improving the user experience and promoting efficient information retrieval. Authors propose in \cite{raj2023generative} a generative model neural conversation system, in order to remedy the limitation of traditional approaches, in terms of domain specifications, and manual rules creation requirements. This system uses a deep LSTM Sequence to Sequence model with an attention mechanism, which aims to address the limitations of previous rule-based approaches by allowing for open domain conversations. This model is trained on Reddit conversation datasets and evaluates its performance using the Turing test. This article presents an interesting approach to building a generative model chatbot using deep learning techniques. The proposed model shows advantageous results and could be further enhanced by incorporating more advanced techniques. The article provides valuable insights into the challenges and opportunities of building chatbots that can have meaningful conversations with humans. Authors in \cite{lhasiw2021bidirectional} developed for the office of the registrar, Thammasat University, a chatbot to answer students' questions. In the context of a chatbot system, accurately identifying the intention behind a question message is a critical task. To dive into this challenge, the article suggests the use of a bidirectional LSTM model for classifying question messages into five different intention classes. The model is qualified and evaluated on a validation dataset, resulting in an accuracy of 0.80. This demonstrates the effectiveness of the bidirectional LSTM approach in accurately categorizing question messages and picking their underlying intentions. The  Frequently Asked Question (FAQ) chatbot implemented in \cite{muangkammuen2018automated} answers customers' questions automatically, using RNN (Recurrent Neural Network), in the form of Long Short-term Memory (LSTM) for text classification. The model proposed goes through these steps: 1) preparing data, 2) pre-processing, 3) classification model, 4) data splitting, and 5) hyperparameters tuning. This model was tested with (2636 pairs de questions/responses), and the results showed that the chatbot could recognize 86,36\% of questions and answers with 93,2\% accuracy. This model is retrieval-based, which learns to select a suitable response from a set of predefined responses,  and unfortunately can not be autonomous and produce answers for new questions. Whereas the generative model could be an alternative in order to improve it since it is able to generate a new response from scratch. Table \ref{tab4} depicts an outline of a generative Chatbot for FAQs.

\begin{center}
\scriptsize
\begin{longtable}[!t]{
m{1.1cm}
m{1.2cm}
m{1.8cm}
m{1.5cm}
m{3.2cm}
m{5.1cm}}
\caption{Summary of generative Chatbot for FAQs.}

\label{tab4}
\\ \hline
 Work & Model & Dataset & Application & Performance & Advantage/Limitation   \\ \hline
\endfirsthead

\multicolumn{6}{c}{{Table \thetable\ (Continue)}} \\
\hline
Work & Model & Dataset & Application & Performance & Advantage/Limitation  \\ \hline 
\endhead

\hline
\endfoot

\cite{serban2017hierarchical} (2017) & RNN &  Reddit conversation datasets  & Social media & Latent variables enhance diverse, meaningful responses.  &   
\newline \textbullet~ Limited scope of evaluation and validation results. 
\newline \textbullet~ Lack of comparison with other dialogue generation models. 
\newline \textbullet~ Presents a hierarchical latent variable encoder-decoder model for dialogue generation. \newline  \\

\cite{huddar2020dexter} (2020) & RNN &  N/A  & Education &   N/A & 
\newline \textbullet~ Limited scope of evaluation and validation results. 
\newline \textbullet~ Lack of specific details on the architecture and techniques used in developing the Dexter chatbot. 
\newline \textbullet~ Proposes insights into the development and implementation of a specific chatbot system tailored for college FAQs. \newline \\

\cite{xu2017new} (2017) & LSTM &  \newline 1M Twitter conversations & Customer service &  > information retrieval system founded &  
\newline \textbullet~ Lack thorough evaluation or user feedback on the chatbot's effectiveness and user experience. 
\newline \textbullet~ Presents a new chatbot especially developed for customer service on social media platforms. \newline \\

\cite{kamphaug2018towards} (2018) & GRU & 200083 questions & Open domain & Accuracy = 71.2\%. &  \newline \textbullet~ Limited coverage of practical implementation challenges, scalability, and real-world performance of the proposed architecture. 
\newline \textbullet~ Presents insights into the development of a chatbot model specially designed for open-domain conversations. \newline \\

\cite{liu2019learning} (2019) & GRU & SQuAD dataset \newline NewQA dataset & Open domain & BLEU-4 = 17.55 \newline ROUGE-L = 44.53 \newline METEOR = 21.24 &  
\newline \textbullet~ Limited discussion on the specific dataset used for training and evaluation in this model. 
\newline \textbullet~ Introduces a novel approach for question generation that focuses on learning what not to generate. \newline  \\

\cite{wang2019answer} (2019) & GRU & Reddit conversation dataset & Open domain & Outperform baseline algorithms in generating coherent. & 
\newline \textbullet~ Limited scalability and generalizability of the proposed approach. 
\newline \textbullet~ Introduces answer-guided and semantic coherent question generation for open-domain conversation. \newline  \\

\cite{adiwardana2020towards} (2020) & Seq2seq &   341GB of text (40B words)  & Social media & SSA = 79\% \newline higher in absolute SSA =  23\% &  
\newline \textbullet~ Ethical considerations 
\newline \textbullet~ Significant progress in the field of open-domain chatbots. \newline \\

\cite{boussakssou2022chatbot} (2022) & Seq2seq & ~81,659 pairs of Q\&A  & Arabic support & N/A & 
\newline \textbullet~ Address languages other than English and French. \newline \\

\cite{khin2020question} (2020) & Seq2seq & N/A & Education &  BLEU = 0.41. &  
\newline \textbullet~ Lack of Comparative Analysis. 
\newline \textbullet~ Presents the first university chatbot implemented in the Myanmar language using a neural network model. \newline  \\

\cite{ricciardelli2019self} (2019) & Deep Q Network &  5 intents & Open domain & 50\% to 75\% success in 20-30 training periods. &   
\newline \textbullet~ Does not explore scalability. 
\newline \textbullet~ Demonstrated that chatbots trained with reinforcement learning show progress in response quality. \newline \\

\cite{xu2022elinor} (2022) & StoryBot & 20 children & Education &  Children's response = 92.8\% \newline Intent classification accuracy = 89\% \newline Children perception = 3 - 4 ( 4 very positive)  &   
\newline \textbullet~ Sample size \& generalizability. 
\newline \textbullet~ Lack of comparison. 
\newline \textbullet~ Real-word application. 
\newline \textbullet~ Positive impact on engagement and learning.\newline \\

\cite{bailey2021digital} (2021) & StoryBot & 27 students & Education & Storybots in L2 class enhance diverse participation; students read 9x more, emphasizing interactive reading's importance. &  
\newline \textbullet~ Limited generalizability. 
\newline \textbullet~ Lack of comparison. 
\newline \textbullet~ Exploration of digital storytelling with chatbots. 
\newline \textbullet~ Contribution to the domain of interactive technology and smart teaching.\newline  \\

\cite{nugraha2022chatbot} (2022) & BoW movies &  Kaggle with a total of 4803 movie data & Entertainment &  Inquiry response rate = 9.1\%.  & 
\newline \textbullet~ Biases in recommendations due to limited movie data and dataset biases. 
\newline \textbullet~ Efficient movie discovery. \newline  \\

\cite{bao2020plato} (2020) & BoW movies &  8M samples & Entertainment &  PLATO-2 surpasses state-of-the-art methods in Chinese and English evaluations.  & 
\newline \textbullet~ Uncertain applicability to other contexts. 
\newline \textbullet~ Multilingual capacity (Chinese and English).\newline \\

\cite{raj2023generative} (2023) & LSTM &  Reddit conversation datasets & Open domain & N/A &  
\newline \textbullet~ Detailed data preprocessing steps, prototype architecture, and training procedure. \newline \\

\cite{lhasiw2021bidirectional} (2021) & LSTM & N/A & Education  & Accuracy = 0,80 &  
\newline \textbullet~ Lack of exhaustive evaluation results or comparisons with other message classification techniques. 
\newline \textbullet~ Presents insights into a specific model architecture for classifying chatbot messages. \newline \\

\cite{muangkammuen2018automated} (2018) & LSTM & 2,636 pairs of Q\&A & Customer service  & Recognition of the questions = 86.36\% \newline Accuracy = 93.2\% & 
\newline \textbullet~ Specific domain. 
\newline \textbullet~ Small dataset for the training process. 
\newline \textbullet~ Address languages other than English and French.  \\

\end{longtable}


\end{center}

\subsection{Hybrid chatbots}
Hybrid chatbots employ a mixture of retrieval-based and generative approaches to effectively address frequently asked questions (FAQs). By integrating these two models, these chatbots can offer precise and contextually appropriate answers to user inquiries. In a hybrid FAQ chatbot, retrieval-based techniques carry primacy in addressing frequently asked questions (FAQs). The chatbot is initially taught on a dataset or FAQ database that contains a range of question-answer pairs. When a user asks a question, the chatbot corresponds the input query with the current questions in its dataset and retrieves the suitable answer. This retrieval method generally employs criteria such as keyword matching or vector similarity to determine the best-matching question and retrieve the corresponding answer. however, in cases where the user asks a question that is not known in the dataset or when the retrieval-based approach fails to find a suitable answer, the hybrid FAQ chatbot can use a generative approach as a fallback. By utilizing a generative model, such as a language model or seq2seq model, the chatbot will be able to generate a response from the ground up using the input question. This capability helps the chatbot address more specific, and complicated queries that may not be covered within the FAQ database, offering a wider scope of assistance to the user. Authors in \cite{da2019ibm} developed a chatbot that engages with students through text messages, around subjects about a Higher Education Institution Course. This chatbot includes the integration of IBM Watson's question-answering capacities with the Moodle platform.  The chatbot was evaluated by facing a set of questions from the same context that were previously responded wrongly. The ultimate aim of this study is to contribute to the development of automated assistance solutions in the learning environment. The authors also underline the possible advantages of using IBM Watson as an FAQ assistant in educational settings, such as decreasing the workload of tutors and enhancing the user experience for users, and they finally examine the limitations such as including the necessity for constant updates and advancements to keep up with growing user queries. In \cite{vu2021online} researchers present a FAQ chatbot for online customer support. They underline the significance of efficient customer support in different domains and how chatbots are practical in automating this procedure. The implemented chatbot consists of several key components. Firstly, there is a user interface that permits users to input their queries and view the system's responses. Secondly, there is a FAQs agent for handling language understanding and dialog management, that was implemented on the DialogFlow platform. Thirdly, there is a web service that serves diverse backend tasks, such as retrieving information from a database and exploring predefined websites using Google Custom Search. In expansion to these, the system includes additional supported modules, such as a crawler to collect data from the school website. Moreover, to improve the performance of the chatbot, they include two sub-components, a module for sentence similarity, to select the most suitable response for a given query, and a question generation module, to generate extra training data for the chatbot agent. While the paper offers valuable insights into the development and implementation process of an online FAQ chatbot for customers, there is room for additional questioning and enhancement. Particularly, on scalability and adaptability, guaranteeing that the chatbot solution can take a larger volume of inquiries and effectively adapt to growing customer needs. 

To show the impact of Deep Reinforcement Learning Chatbot, authors in \cite{serban2017deep} present MILABOT, a chatbot produced by the Montreal Institute for Learning Algorithms (MILA) for the Amazon Alexa Prize competition. This chatbot was developed to engage in interactions with users on various popular topics, using both speech and text modalities. The system is hybrid and combines natural language generation (NLG) and retrieval models,sequence-to-sequence, and latent variable neural network models. The development of MILABOT takes advantage of reinforcement learning techniques on crowdsourced data and real-world user interactions. This chatbot was evaluated through A/B testing with real users, where MILABOT shows excellent performance compared to multiple competing systems. Authors in \cite{reyes2019methodology} present a chatbot implemented in Google Dialogflow, designed for students to get interactive content. This chatbot presents a hierarchical structure of frequently asked questions, permitting students to navigate through the content, in order to find answers to their specific questions and concerns. The chatbot proposed obtained positive feedback from students, who enjoyed its accessibility and well-structured data.
Table \ref{tab5} presents a comparison of existing hybrid chatbots for FAQs.

\begin{center}
\scriptsize
\begin{longtable}[!t]{
m{1.1cm}
m{1.2cm}
m{1.3cm}
m{1.5cm}
m{3.9cm}
m{4.9cm}}
\caption{Summary of hybrid Chatbot for FAQs.}

\label{tab5}
\\ \hline
 Work & Model & Dataset & Application & Performance & Advantage/Limitation    \\ \hline
\endfirsthead

\multicolumn{6}{c}{{Table \thetable\ (Continue)}} \\
\hline
Work & Model & Dataset & Application & Performance & Advantage/Limitation  \\ \hline 
\endhead

\hline
\endfoot

\cite{da2019ibm} (2019) & \newline Machine learning algorithms \newline Probabilistic models \newline  &  N/A  & Education & N/A & 
\newline \textbullet~ Lack of Dataset Information.
\newline \textbullet~ Scope and Generalizability. 
\newline \textbullet~ Practical Application.
\newline \textbullet~ Reduce instructors workload. \newline\\

\cite{vu2021online} (2021) & NLG \newline RNN \newline Intent and entity recognition \newline & 2600 Q\&A from Askntu   & Customer service & 
\newline \textbullet~ Good response to trained intentions. 
\newline \textbullet~ Struggles with understanding new queries. & 
\newline \textbullet~ Lack of discussion on the specific architecture and techniques used in designing the online FAQ chatbot. 
\newline \textbullet~ Presents an online FAQ chatbot for customer support, handling a practical application of chatbot technology. \newline \\

\cite{serban2017deep} (2017) & NLG \newline Retrieval models \newline Seq2seq & Datasets scraped from Reddit & Open domain (SocialBot) & 
\newline \textbullet~ Top-performing system in semi-finals. 
\newline \textbullet~ Achieved average user score 3.15 on 1-5 scale. \newline  & 
\newline \textbullet~ Supplies insights into the application of deep reinforcement learning methods in chatbot systems. \newline \\

\cite{reyes2019methodology}  (2019) & DialogFlow &  N/A  & Education &
\newline \textbullet~ Students widely approved the final product. 
\newline \textbullet~ Praise for accessibility and organized data. & 
\newline \textbullet~ Lack of Dataset Information.
\newline \textbullet~ Scope and Generalizability. 
\newline \textbullet~ Practical Application.
\newline \textbullet~ Time savings for students.  \\

\end{longtable}


\end{center}

\section{ChatGPT}
ChatGPT, designed by OpenAI, is a very sophisticated chatbot that has garnered considerable attention and glory in recent months. ChatGPT is the successor of the language model GPT (Generative Pre-trained Transformer), which excelled in generating coherent and contextually appropriate answers, ChatGPT particularly concentrates on conversational exchanges, to make it appropriate for chatbot applications and human-computer interactions. The schema of ChatGPT is based on a Transformer model, that uses attention mechanisms to catch dependencies between words and generate high-quality responses. By utilizing large-scale pre-training on diverse and extensive text data, ChatGPT can mimic human conversational patterns. ChatGPT made a significant improvement in the field of generative conversational AI, obtaining a tool close to creating intelligent and interactive chatbot systems. Since it has very notable key features, such as Natural Language Understanding, Contextual Understanding, and Flexibility and Adaptability. As an analysis in this domain, it is believed that this model will continue to progress, and further advancements in chatbot technology and its potential to revolutionize human-computer interactions can be envisioned. Numerous recent publications have focused on chatGPT, in \cite{hulman2023chatgpt}, authors compare chatGPT to human experts on diabetes FAQ. Danish diabetes center staff was surveyed for response evaluation, assessing accuracy, comprehensibility, and perceived expertise. The results revealed that ChatGPT's responses matched human experts in accuracy and comprehensibility but were considered less knowledgeable and trustworthy. Specifically, ChatGPT performed better when trained on diabetes-specific data, emphasizing the importance of domain-specific training for chatbot performance in specific areas. In summary, the article sheds light on ChatGPT's reliability for diabetes info but stresses domain-specific training and ongoing monitoring for safety and quality. In another study \cite{dwivedi2023so} researchers present a multidisciplinary view on the opportunities, challenges, and implications of generative conversational AI, particularly ChatGPT, for research. This article highlights AI chatbots' educational potential for personalized support. Emphasizes ethical considerations and interdisciplinary collaboration for responsible development and deployment.

\subsection{Advantages of ChatGPT in FAQ Chatbots}

ChatGPT, as a generative language model, offers several ways to enhance the capabilities of FAQ chatbots by generating dynamic and contextually appropriate responses. It excels in providing personalized responses based on specific contexts and user preferences, thanks to its natural language generation capabilities that produce human-like responses. With its extensive pre-training on vast text datasets, ChatGPT is adept at delivering precise and contextually relevant answers to frequently asked questions (FAQs). Its ability to maintain conversation context throughout interactions is valuable for FAQ chatbots. Additionally, by integrating ChatGPT with a FAQ dataset, it can acquire structured information and retrieve answers from the database as needed, further enhancing its effectiveness as an FAQ chatbot. In overview, ChatGPT's generative capacities deliver powerful enhancements to FAQ chatbots. It helps them to address unknown questions for the FAQ dataset, manage ambiguity and context, and continuously learn and improve. Blending ChatGPT into the FAQ chatbot can significantly improve the user experience, leading to more interesting and satisfying interactions.

\subsection{NLP Techniques}

NLP is a field of artificial intelligence destined for the interaction between computers and human language. NLP techniques help computers to understand, analyze, and generate meaningful and useful human language. The well-known tool chatGPT uses NLP approaches to apprehend and generate human-like text. ChatGPT incorporates various NLP techniques to enhance its text-understanding capabilities. These techniques include text preprocessing, which involves tasks such as tokenization to break down text into individual words, part-of-speech tagging to assign grammatical tags to each word, and entity name recognition to identify individuals, institutions, places, dates, and more. Additionally, ChatGPT utilizes sentiment analysis to determine the sentiment expressed in a text, whether it is positive, negative, or neutral. Another valuable feature is the ability to generate concise summaries of longer texts. These NLP techniques enable ChatGPT to analyze and process text effectively, facilitating accurate and meaningful interactions. NLP has an important role in numerous applications such as chatbots, virtual assistants, sentiment analysis, and more. It helps computers to comprehend and process human language, enabling natural and intuitive interactions with machines.

\subsubsection{Named entity recognition}
In the context of chatbots, named entity recognition (NER) is a helpful method, that allows them to recognize and extract, from user queries or conversations, specific entities, such as names of individuals, institutions, places, dates, etc. When integrated into chatbots. This approach offers several functionalities that contribute to improved natural language understanding and accuracy. To provide insights into the advancements in NER, the authors in \cite{sun2018overview} present an overview established on the research of 162 papers from NLP conferences, and also expose a summary of results and highlight potential coming trends for NER research. In another study \cite{hu2023zero} that compares ChatGPT, GPT-3, and BioClinicalBERT on clinical named entity recognition, ChatGPT outperformed GPT-3 in zero-shot settings, though still behind the supervised BioClinicalBERT. The results highlight ChatGPT's potential in this field due to its capability of achieving decent results without the need for extensive annotation, enhancing feasibility and cost-effectiveness. To sum up, including NER in chatbot systems consolidates them to apprehend and respond effectively to user queries, providing more precise and personalized interactions. It improves the user experience by helping chatbots to identify and operate with named entities just like human comprehension and communication.

\subsubsection{Intent classification}
Intent classification in NLP systems, like ChatGPT, discerns the underlying user intent, enhancing conversational capabilities and personalizing interactions. For intent classification, ChatGPT is trained on datasets with annotated inputs, helping it identify textual patterns indicative of specific intents. Upon classifying an intent, ChatGPT can offer tailored responses using predefined templates linked to each intent or route queries to the correct system domain. A study \cite{cegin2023chatgpt} explored replacing human crowdsourcing with ChatGPT for paraphrase generation in intent classification dataset creation. The findings revealed that ChatGPT-generated paraphrases were more diverse and equally robust as human-sourced ones but at a lower cost. While effects on model robustness varied based on the data source, merging both human and ChatGPT data could optimize model performance, highlighting ChatGPT's cost-effective data collection advantage.

\subsubsection{Sentiment analysis}
Sentiment analysis, a key technique in NLP, identifies emotional tones in text, finding use in areas like social media monitoring and brand reputation management. While ChatGPT doesn't inherently have sentiment analysis features, existing methods or specialized models can be employed to assess the sentiment of its outputs \cite{essop2023developing}. These models categorize text as positive, negative, or neutral. However, analyzing sentiment in ChatGPT's outputs presents challenges. Since ChatGPT generates responses based on statistical patterns, it might not consistently exhibit clear sentiment. The efficacy of sentiment analysis also heavily relies on training data quality and the selected approach. In essence, while it's possible to apply sentiment analysis to ChatGPT's outputs, there are inherent limitations and challenges to consider \cite{alemdag2023effect}.

\subsection{Training and Fine-tuning ChatGPT for FAQ Chatbots}
Training ChatGPT using FAQ datasets is a strategic process aimed at enhancing the model's capability to provide contextually accurate information. The training journey begins with dataset preparation, selecting a particular FAQ set, followed by pretraining on a vast internet text corpus for foundational language skills \cite{ge2023designing}. Fine-tuning is then carried out using the chosen FAQ dataset, ensuring the model adapts to the specific language and style, while an iterative feedback loop with users and experts refines its responses. The essence of this methodological approach underscores the paramount importance of high-quality training data. A robust dataset not only influences ChatGPT's performance in delivering reliable answers but also equips it to recognize diverse linguistic patterns and handle various question phrasings. Moreover, curated and validated data minimizes biases, ensuring the chatbot's answers remain both trustworthy and objective \cite{ait2023impact}.

\subsection{Integration with Knowledge Base and Retrieval Mechanisms}
Integrating ChatGPT with a knowledge base or retrieval mechanisms enhances its ability to offer precise, relevant answers. By tapping into structured information or FAQs within a knowledge base, ChatGPT can pull specific, accurate answers, especially when combined with real-time information retrieval from sources like search engines. Training the model using both conversational data and the knowledge base content ensures it delivers contextually apt answers. Ensuring the model's contextual understanding further refines its performance \cite{balderas2023chatbot}. To maintain this system's efficacy, the knowledge base must be continually updated and expanded. Algorithms that rank and filter answers can help ChatGPT select the most pertinent responses from the knowledge base. Keeping the knowledge base updated is pivotal for ChatGPT's accuracy and user satisfaction. Yet, challenges such as data collection, scalability, and real-time updates arise when syncing the knowledge base with ChatGPT. Overcoming these hurdles requires efficient strategies, robust systems, and continuous monitoring, ensuring the knowledge base meets users' expectations and boosts their engagement \cite{chen2023adoption}.

\subsection{Customization and Personalization}
ChatGPT offers a remarkable advantage when it comes to implementing FAQ chatbots in organizations or websites. It is able to customize and tailor the chatbot to specific needs and branding. This grade of customization permits organizations to create a chatbot experience that aligns with their identity, providing personalized and cohesive interaction for users. By exploring the flexibility of ChatGPT, organizations can develop a chatbot that reflects their brand, provides accurate and accurate responses, integrates with existing systems, and presents valuable insights. Table \ref{tab6} describes the various aspects of customization that make ChatGPT an exceptional choice for organizations seeking a unique and branded FAQ chatbot experience.

\begin{center}
\scriptsize
\begin{longtable}{
m{1.7cm}
m{5.5cm}
m{2.5cm}
m{2.5cm}
m{2.5cm}}
\caption{Aspects of Customization in ChatGPT for a Unique and Branded FAQ Chatbot Experience.}
\label{tab6}
\\ \hline
Aspect & Description & Examples & Benefits & Challenges \\ \hline
\endfirsthead

\multicolumn{5}{c}{{Table \thetable\ (Continue)}} \\
\hline
Aspect & Description & Examples & Benefits & Challenges \\ \hline
\endhead

\hline
\endfoot

Branding Alignment & 
Customize to reflect the organization's brand identity, incorporate logos, colors, and visual elements, and create a seamless brand experience for users interacting with ChatGPT.
& Company colors in ChatGPT's UI; Company logo in chat window
& Increases brand recognition; Enhances trust
& Over-customization; Maintaining brand consistency

\\

Specific FAQ Content & 
Train ChatGPT with organization-specific FAQs, enable accurate responses to frequently asked questions, and customize ChatGPT's knowledge of the organization's unique information.
& Integrating industry-specific jargon; Organization's product FAQs
& Relevant and tailored user experience; Reduces user search time
& Continuous updates needed; Frequent re-training for accuracy

\\

Customized Responses & 
Use ChatGPT to assist in defining and tailoring responses, customize responses using ChatGPT for accuracy and usefulness, and allow organizations to create adapted responses with ChatGPT.
& Setting up response templates; Guided responses
& Personalized user experience; Increases user satisfaction
& Balance between automation and personal touch; Over-customization

\\

Language and Tone Customization & 
Adapt ChatGPT's responses to match the organization's communication style, customize ChatGPT's answers to reflect the organization's brand voice, and ensure consistency in communication by aligning ChatGPT's responses.
& Adapting a formal tone for a law firm; Casual language for a youth brand
& Resonates with target audience; Aligns with brand voice
& Maintaining tone consistency; Avoiding inappropriate tone

\\

Smooth Integration & 
Smooth integration of ChatGPT with current systems and databases, real-time information access facilitated by using ChatGPT, and personalized responses by incorporating existing data with ChatGPT.
& Integration with CRM systems; Syncing with databases
& Real-time data access; Seamless user experience
& Compatibility issues; Ensuring data security

\\

Analytics and Insights &
Provide analytics and insights through ChatGPT on user interactions, enable data gathering on user behavior using ChatGPT's capabilities, and deliver valuable insights to organizations from user exchanges with ChatGPT.
& Tracking user interaction duration; Monitoring frequent questions
& Provides actionable insights; Understanding user behavior
& Data overload; Ensuring user privacy

\\

Multi-lingual Support & 
Personalize ChatGPT for supporting multiple languages, enhance global accessibility with ChatGPT's multi-language capability and expand the user base by enabling interactions in different languages using ChatGPT.
& Offering ChatGPT support in Spanish; FAQ support in local languages
& Enhances global reach; Improved satisfaction in non-English regions
& Ensuring accurate translation; Maintaining context in languages

\\

Sentiment Analysis & 
Integrate emotion analysis capabilities into ChatGPT's customization, customize ChatGPT to recognize and respond to user emotions effectively, and enhance ChatGPT's responsiveness by incorporating emotion analysis.
& Detecting frustration through repeated questions; Recognizing feedback through emojis
& Better understanding of user sentiment; Proactive response
& Misinterpretation of sentiment; Over-relying on automated sentiment

\\

Adaptive User Experience & 
Customize the potential of ChatGPT for adaptive user experiences, dynamic response adjustments by ChatGPT based on user behavior, and personalize recommendations by ChatGPT according to user preferences.
& Adjusting responses based on browsing history; Personalizing suggestions
& Enhanced engagement; Tailored user experience
& Balancing personalization and privacy; Avoiding perceived invasiveness

\\

Advanced Search Capabilities & 
Harness the advanced search functionalities of ChatGPT, enable users to perform targeted searches within ChatGPT's knowledge base, and use ChatGPT for efficient information retrieval through advanced search.
& Enabling keyword-based searches; Searching within archived chats
& Faster information retrieval; Improved user experience
& Ensuring search accuracy; Handling misinformation

\\

\end{longtable}
\end{center}

Personalizing responses based on user preferences or previous interactions holds great potential for enhancing the user experience. By gathering data and insights from user interactions, ChatGPT can personalize its responses to align with individual preferences and specific needs. This includes delivering customized recommendations, adjusting the conversation flow based on earlier interactions, considering the context for more pertinent responses, employing user-specific information, adjusting language and communication style, providing proactive assistance, and recognizing and responding to user emotions. By personalizing responses, ChatGPT creates a more engaging and smooth experience, thus boosting user satisfaction, trust, and loyalty \cite{omar2023chatgpt}.

\subsection{Limitations and Ethical Considerations}


ChatGPT, while beneficial for FAQ chatbots, presents ethical concerns due to its reliance on extensive online text data, which can lead to inaccuracies and biases. Small input variations may result in unreliable answers, potentially provoking user confusion. Furthermore, it can exhibit undue confidence in responses without indicating uncertainty levels. A summarized overview of these limitations is provided in Table \ref{tab7} to guide responsible ChatGPT implementation in FAQ systems. Addressing these ethical considerations requires a commitment to ethical guidelines and responsible AI practices. Developers and organizations deploying chatbots like ChatGPT need to prioritize transparency, responsibility, and justice in their design and implementation. Standard monitoring, auditing, and user feedback mechanisms can help specify potential ethical issues, ensuring that the technology is employed in a healthful and responsible way.

\begin{center}
\scriptsize
\begin{longtable}[!t]{
m{2.5cm}
m{7.5cm}
m{1.5cm}
m{3.3cm}}
\caption{ChatGPT limitations and ethical considerations.}
\label{tab7}
\\ \hline
Limitation & Description & Reference & Eventual solution  \\ \hline
\endfirsthead
\multicolumn{4}{c}{{Table \thetable\ (Continue)}}  \\ \hline
Limitation & Description & Reference & Eventual solution  \\ \hline 
\endhead
\\ \hline
\endfoot
Lack of factual accuracy  & 
\newline \textbullet~ ChatGPT's knowledge relies on existing data that might not be up-to-date or trustworthy. 
\newline \textbullet~ It learns from extensive internet text data, which can include subjective or incorrect information. 
\newline \textbullet~ As a result, ChatGPT may provide responses that are inaccurate or nonobjective. &  \cite{hariri2023unlocking} (2023) \newline \cite{limna2023use} (2023) \newline \cite{biswas2023role} (2023) \newline \cite{sallam2023chatgpt} (2023) & 
\newline \textbullet~ Monitoring the training data  \\

sensitive to small modifications in users’ & 
\newline \textbullet~ Susceptible to responding acutely to minor alterations in user inputs. 
\newline \textbullet~ Exhibiting inconsistency or randomness in its responses. 
\newline \textbullet~ This behavior can lead to user confusion and frustration. &  N/A  & 
\newline \textbullet~ Robust training Data. 
\newline \textbullet~ Controlled Generation. 
\newline \textbullet~ Post-Processing and Filtering. 
\newline \textbullet~ User Feedback and iterative improvement.   \\

overconfidence & 
\newline \textbullet~ ChatGPT shows overconfidence through highly certain responses. 
\newline \textbullet~ Provide definitive answers even with insufficient supporting information. 
\newline \textbullet~ Potential for users to be misled by overly confident, unreliable information.
\newline \textbullet~ Confident ChatGPT responses can lead to accepting wrong information. \newline &  N/A   & 
\newline \textbullet~ Ongoing Monitoring and Evaluation and thus Confidence Adjustment  \\

Verbosity and overuse of certain phrases & 
\newline \textbullet~ It is excessive word usage or unnecessary detail in communication. 
\newline \textbullet~ Can lead to redundancy and lack of conciseness in expression. 
\newline \textbullet~ Overusing specific phrases can dilute message impact and create monotony.\newline &  \cite{ray2023chatgpt} (2023)  & 
\newline \textbullet~ Fine-tuning.
\newline \textbullet~ Human Review and Feedback. 
\newline \textbullet~ Reinforcement Learning.
\newline \textbullet~ Response Filtering. \\

difficulties in recognizing and adapting to user expertise & 
\newline \textbullet~ Recognize challenges in adapting to user expertise for models like ChatGPT.
\newline \textbullet~ Confronte difficulty in tailoring responses to user expertise. 
\newline \textbullet~ ChatGPT's responses are rooted in training data patterns, not user-specific expertise. 
\newline \textbullet~ Limitation of ChatGPT in considering user's proficiency or background. &  \cite{ray2023chatgpt} (2023)  & 
\newline \textbullet~ User Profiling. 
\newline \textbullet~ Feedback and Iterative Improvement. \\

Inconsistency in response quality & 
\newline \textbullet~ Acknowledge response quality inconsistency challenge in models like ChatGPT.
\newline \textbullet~ Context gaps, query ambiguity, training biases, and data noise contribute to the issue. &  \cite{ray2023chatgpt} (2023) \newline \cite{hariri2023unlocking} (2023) & 
\newline \textbullet~ Improve model training.
\newline \textbullet~ Enhance contextual understanding. 
\newline \textbullet~ Implement reinforcement learning and feedback loops.
\newline \textbullet~ Explore fine-tuning and domain-specific training.  \newline\\

Overgeneralization & 
\newline \textbullet~ Overgeneralization is a prevalent challenge in language models like ChatGPT.
\newline \textbullet~ Manifests through overly broad or inaccurate statements not suiting specific cases. 
\newline \textbullet~ Originating from biased training data, lack of real-time data, and inadequate context. 
\newline \textbullet~ Noise in the data also contributes to the issue of overgeneralization. &  \cite{ray2023chatgpt} (2023)  &  
\newline \textbullet~ Fine-tuning and domain-specific training.
\newline \textbullet~ Data curation, and bias handling. 
\newline \textbullet~ Improve contextual understanding.
\newline \textbullet~ Incorporate user feedback for iterative improvement.  \newline\\

Data privacy and security & 
\newline \textbullet~ Highlight the importance of safeguarding user privacy in chatbots like ChatGPT.
\newline \textbullet~ Emphasize the need for adequate measures to protect user data. 
\newline \textbullet~ Ensure user information security through appropriate precautions. 
\newline \textbullet~ Acknowledge that ChatGPT's data dependence necessitates privacy protection.
\newline \textbullet~ Underscore the significance of implementing safeguards for user privacy. \newline &  \cite{ray2023chatgpt} (2023) \newline \cite{limna2023use} (2023) \newline \cite{sallam2023chatgpt} (2023) &  
\newline \textbullet~ Data encryption. 
\newline \textbullet~ Anonymization. 
\newline \textbullet~ Regular audits and updates. \newline  \\

Intellectual property and authorship & 
\newline \textbullet~ ChatGPT's data-driven responses raise content ownership concerns. 
\newline \textbullet~ Generating questions about the source and ownership of the content it creates. \newline &  \cite{ray2023chatgpt} (2023)  &  
\newline \textbullet~ Set guidelines to safeguard content creators' rights.  \\

Transparency and explainability & 
\newline \textbullet~ Transparency in ChatGPT's decision-making might not be constant for users.
\newline \textbullet~ Emphasizing the significance of providing reasons for generated responses.
\newline \textbullet~ Improving user trust and system comprehension through transparent explanations. \newline &  \cite{ray2023chatgpt} (2023)  &  
\newline \textbullet~ User-Friendly Explanations.
\newline \textbullet~ Interpretable Outputs. \\

Environmental impact & 
\newline \textbullet~ Large model training consumes resources, impacting the environment. 
\newline \textbullet~ Exploring energy-efficient methods or reducing environmental impact is vital.
\newline \textbullet~ Emphasizing addressing bias and discrimination as ethical imperatives. &  \cite{ray2023chatgpt} (2023)  & 
\newline \textbullet~ Explore options for supporting environmental offsets. \newline \\

Responsibility and accountability  & 
\newline \textbullet~ Establish accountability for chatbot development and deployment. 
\newline \textbullet~ Involves ethics, risk reduction, and managing unintended outcomes. &  \cite{ray2023chatgpt} (2023)  &  
\newline \textbullet~ Clear Guidelines and Policies.
\newline \textbullet~ Transparent Documentation. 
\newline \textbullet~ User Feedback and Reporting.
\newline \textbullet~ Audits and Evaluations.
\newline \textbullet~ Collaboration with Experts. \\

\end{longtable}

\end{center}

\section{Applications}
ChatGPT's adaptability makes it a valuable tool across numerous industries. In online shopping, it improves the customer experience by answering product queries and offering recommendations. In customer service, it swiftly resolves issues, while in content creation, it aids in drafting and brainstorming. As a digital tutor in education, it provides clarifications and examples, and in gaming, it acts as an interactive character. ChatGPT offers support in healthcare by providing pertinent information, assisting with financial inquiries in finance, facilitating trip planning in travel and hospitality, delivering news updates in media, aiding research tasks, and serving as a virtual HR assistant. This versatility underscores ChatGPT's transformative potential across diverse domains.


\subsection{Online shopping}
ChatGPT, OpenAI's advanced language model, has gained traction in various sectors, notably online shopping with personalized experiences, real-time support, and 24/7 availability. Advantages include tailored product recommendations and boosted user engagement. However, it may struggle with difficult questions and lacks real-time updates, needing data integration for optimal performance. Balancing its advantages and limitations is key to a reliable shopping experience. Many publications explore ChatGPT's applications, Authors in \cite{hariri2023unlocking} suggest using ChatGPT to create customer service chatbots. These chatbots, powered by ChatGPT's understanding of natural language, can personalize responses, thereby enhancing user experiences and lightening the load for customer service representatives. In a study by \cite{george2023review}, authors spotlight ChatGPT's e-commerce potential. It cuts costs by automating customer queries, reduces human support dependence, and aids scalability during peak periods, ensuring timely responses and operational efficiency. Thus, integrating ChatGPT offers businesses cost efficiency and superior operational capability, by both cutting down human support costs and effectively handling peak traffic periods. In a separate research \cite{paul2023chatgpt}, authors outline the advantages of using ChatGPT to enhance customer experiences, by Exploiting advanced language processing for personalization and real-time communication. Economically, businesses can benefit from the cost-effectiveness of ChatGPT in customer service, as automating some support functions reduces reliance on human agents, leading to efficient resource allocation and savings. Furthermore, by analyzing customer interactions, ChatGPT can refine marketing strategies, delivering targeted and personalized campaigns, thus augmenting the impact of promotional efforts. Furthermore, the incorporation of ChatGPT into online shopping presents myriad advantages. It significantly elevates the online shopping experience by offering tailored product suggestions in line with customer preferences. As a proficient virtual shopping assistant, ChatGPT adeptly handles customer inquiries, supplies comprehensive product details, and responds to common queries. It supports customers throughout their shopping experience, from product selection to checkout. Valuably, it also extracts insights from customer feedback to refine offerings. Beyond product specifics, ChatGPT assists with size and color choices, aids in comparing products, and guides users on payment options. It also efficiently manages product returns and exchanges, guaranteeing a seamless shopping journey. By promoting special deals and discounts, ChatGPT further enriches customer engagement and drives sales. 

\subsection{Healthcare}
FAQ chatbots, leveraging AI and NLP, have transformed healthcare by offering immediate and accurate answers to common queries, enhancing patient engagement, supporting self-care, and reducing pressure on healthcare professionals. Subsequently, numerous studies have explored this domain. In the study \cite{biswas2023role}, the authors outline multiple ways ChatGPT can be beneficial in healthcare. It can (i) disseminate timely information on public health matters, such as disease outbreaks and vaccinations, (ii) Offer guidance on health promotion and disease prevention, including personalized advice on maintaining a healthy lifestyle and managing chronic illnesses, (iii) explain the roles of community health workers and educators, highlighting their contributions, (iv) examine how social and environmental factors influence community health, helping to identify trends and possible interventions, and (v) share details about community health programs and local healthcare resources. These applications underscore ChatGPT's potential to enhance health information distribution, promote preventive health, assist community health efforts, and boost healthcare accessibility and knowledge. The study by \cite{li2023chatgpt} offers a comprehensive exploration of ChatGPT's role in healthcare. Authors categorize ChatGPT literature into application and user-focused areas in healthcare. In application-oriented contexts, ChatGPT sees use in triage, translation, research, workflow optimization, education, consultation, and multimodal applications. User-oriented beneficiaries include patients, professionals, researchers, students, and educators. The authors also weigh the technology's strengths, like real-time responses, and note limitations such as privacy concerns and biases, emphasizing responsible AI integration in healthcare. In another work \cite{cascella2023evaluating}, the authors examine the feasibility and implications of using ChatGPT in healthcare, a critical sector impacting human health. They investigate its applications in clinical practice, research, potential misuse, and public health discussions. Emphasizing the need for responsible usage, the authors advocate for education on the technology's capabilities and limitations. This knowledge empowers stakeholders to make informed choices, balance potential risks, and maximize the benefits of AI chatbots. The study accentuates the importance of integrating AI in healthcare responsibly, with an emphasis on ethics and patient well-being. Another researcher who also was interested in ChatGPT, David A. Asch, MD, a senior figure at the University of Pennsylvania, interviewed ChatGPT, examining its potential in healthcare \cite{asch2023interview}. While acknowledging ChatGPT's impressive conversational abilities, David A. expressed concerns about the tool's accuracy and reliability in delivering medical information. To conclude, in this study \cite{javaid2023chatgpt}, the authors outline various applications of ChatGPT in healthcare as depicted in Figure \ref{fig:taxonomyG}. However, they caution against relying solely on ChatGPT for medical advice, emphasizing it shouldn't replace medical professionals. Notably, ChatGPT's knowledge only extends to September 2021, potentially missing recent medical developments. To enhance its utility in healthcare, refining ChatGPT with medical datasets, updating its knowledge, and incorporating continuous feedback are vital steps.


\subsection{Education}
AI-powered FAQ chatbots enhance education, providing rapid answers, improving student experiences, and simplifying administrative tasks. In \cite{tlili2023if}, authors explore ChatGPT's educational potential, emphasizing personalized, interactive learning. ChatGPT engages students, answers questions, and promotes active learning. Challenges include potential inaccuracies due to reliance on existing data and alignment with educational goals and pedagogical principles for effective use. In another paper \cite{sok2023chatgpt}, researchers examine the advantages and disadvantages of employing ChatGPT in education. ChatGPT can enhance learning with interactive assessments and personalized tutoring but raises concerns about academic integrity and misinformation. A proportional approach implicates promoting critical thinking, technology awareness, and using ChatGPT alongside human educators for maximum benefits and risk mitigation. The article by \cite{opara2023chatgpt} delves into considerations and challenges of using ChatGPT in education. Key concerns include over-reliance on ChatGPT, potentially sidelining human guidance and expertise vital for effective learning. Another problem is the lack of citations for ChatGPT's outputs, leading to plagiarism risks. Thus, students must be educated about proper referencing. The potential for ChatGPT to provide incorrect information is also a concern, emphasizing the need for critical evaluation of its responses. Additionally, ChatGPT might offer limited answers to complex queries, impacting the depth of learning. The authors stress the importance of a balanced approach, blending human involvement and leveraging ChatGPT while staying aware of its limitations. This insight aids educators in optimizing ChatGPT's role in the classroom. Moreover, ChatGPT is reshaping education, offering multiple benefits outlined in Figure \ref{fig:taxonomyG}, which shows a list of how ChatGPT can be used in the field of education. ChatGPT can act as an immediate resource for students, offering clarifications like a virtual tutor. It assists with homework, gives in-depth explanations, and supports language practice and exam preparation. It fosters critical thinking, provides writing feedback, and aids in study guidance. Additionally, it enhances revision by providing examples, offers language translation, and encourages group collaborations. For teachers, it streamlines classroom management, letting them concentrate on teaching. It also offers career advice and supports virtual classrooms, enriching students' learning. In essence, ChatGPT enriches education, heightening engagement and personalizing learning experiences.



\section{Open chanllenges}
Many studies are being conducted to set up an irreproachable chatbot, which will be able to be autonomous, and thus provide coherent answers. Based on the literature review made in the previous section, this section includes some chatbot drawbacks that can alter its effectiveness. Many chatbot limitations are presented in \cite{ayanouz2020smart}, such as Grammatical Errors, Predefined or Closed-domain, Language Structure, etc.

\begin{itemize}
\item \textbf{Context problem and the semantic:} chatbot is a robot that can determine the context of a sentence if it is not defined properly by its interlocutor. A chatbot can also face a lot of semantic issues, depending on the language used, since a word can have many several senses, depending on the sentence context.

\item \textbf{Grammar mistakes and structure of the language:} Most of the existing chatbots are based in one language at the same time since the rules from one language to another change constantly. Moreover, questions asked by users can sometimes be ambiguous due to grammatical issues or may require additional context to answer correctly. Designing a model that can ask clarifying questions or manage ambiguity in a user-friendly manner is an important research challenge.

\item \textbf{Accuracy rate:} The answers provided by a chatbot must be related to the questions, or at least close to the context of the question. What would be the use of a chatbot that always answers questions, randomly, but that comes out of the context of the question that was asked? It is not enough for a chatbot to answer questions automatically and autonomously, but also to align itself with the requests of its interlocutors.

\item \textbf{Analysis of Emotions:} In order to bring the chatbot closer to the human, the conversation inherits human characteristics. It is essential to integrate the analysis of sentiments into the conversations provided by chatbots, and thus maintain the conversation longer with the interlocutor. Authors in \cite{zhou2020design} recommend that a social chatbot needs to present a consistent personality to set the right expectations for users in the conversation and attain their long-term confidence and trust. 

\item \textbf{Self-learning:} Machine learning-based chatbots can become more and more autonomous, as they are able to enrich their knowledge base through exchanges, and so these systems will be in continuous progression. Furthermore, uncontrolled machine learning techniques can cause considerable damage, such as that encountered by the TAY Microsft chatbot. This chatbot caused subsequent controversy when it began to post inflammatory and offensive tweets through its Twitter account, which made Microsoft shut down the service only 16 hours after its launch\cite{chhabra2021overview}.

\item \textbf{Evaluation Metrics:} Current evaluation metrics for generative chatbots (like BLEU, ROUGE, etc.) do not always correlate well with human judgment. Developing better evaluation metrics for such chatbots is an open research problem.

\item \textbf{Interpretability:} Generative models, especially deep learning-based ones, are often considered black boxes, i.e., their decision-making process is hard to understand. Making these models interpretable is an open research problem.

\end{itemize}

In order to remedy the above problems, a chatbot should be developed considering human capabilities, since it should be able to understand the semantics and the context of the sentences, analyze the emotions of its interlocutor, and then respond with a high accuracy rate.

\section{Future Research Directions}

ChatGPT's introduction has significantly advanced FAQ chatbots, paving the way for better user interactions and conversational experiences. This section delves into potential research areas, highlighting improvements in conversational flow, multi-turn interactions, and integrating ChatGPT with other AI technologies. It also touches upon addressing biases, and ethical considerations, and offering user-centric customizations. Pushing ChatGPT's boundaries can revolutionize FAQ chatbots, fostering more intelligent and engaging dialogues.

\subsection{Optimizing conversational flow}
Permanent advancements in Natural Language Understanding (NLU) can revolutionize the capacities of ChatGPT. Using improved NLU, ChatGPT models can accurately understand the nuanced meaning of user queries, accurately extract pertinent information, and generate more meticulous and contextually suitable responses \cite{singh2023chatgpt}. Furthermore, the integration of contextual understanding within ChatGPT models can enable them to maintain and utilize information from previous turns in a conversation. This context retention improves the coherence of the conversation, permitting ChatGPT to provide more contextually relevant and seamless responses throughout the conversation. User understanding and personalized adaptation is another favorable research area that can contribute to personalized interactions. ChatGPT can adjust its responses and conversation style to align with the user's needs and preferences by developing a deeper knowledge of user preferences, personal characteristics, and conversational history. This personalized approach creates a more interesting and satisfying conversational experience for the user. To conclude, these improvements in NLU, contextual understanding, user understanding, and personalized adaptation can collectively boost the conversational abilities of ChatGPT, making it more precise, legible, and personalized in its interactions with users \cite{singh2023leveraging}.

\subsection{Advancing dialogue across multiple turns}
Future iterations of ChatGPT are predicted to show improved memory retention for better recall abilities, allowing them to remember and recover information from earlier turns in a conversation. This enhanced memory retention authorizes ChatGPT to provide more significant and engaging multi-turn exchanges by including suitable context and building according to previous exchanges. Moreover, ChatGPT could evolve to incorporate advanced conversation planning mechanisms. By generating coherent and goal-oriented conversation plans, the model can strategically structure the flow of dialogue, ensuring smoother transitions between turns and enabling more effective management of the conversation dynamics. This enhances the overall conversational experience and makes interactions with ChatGPT feel more natural and purposeful. When evaluating ChatGPT's performance in an exam scenario, a series of exam-related questions were posed. To counteract memory retention bias, a fresh chat session was conducted for each question. Recurrent neural networks can lead to memory retention, where learned information influences subsequent data inputs and outputs \cite{lum2022can}. To handle unclear or incomplete user queries, ChatGPT could utilize methods for question clarification \cite{loni2011survey}. By asking follow-up questions to collect additional information, the model can overpower ambiguities and fill in gaps in user input. This helps ChatGPT to provide more accurate and relevant responses, enhancing the quality of the interaction and user satisfaction. These improvements in memory retention for better recall, conversation planning, and question clarification collectively contribute to the evolution of ChatGPT, allowing it to engage in more refined, contextually aware, and effective multi-turn conversations.

\subsection{Improving transfer learning and domain adaptation}
Transfer learning, a facet of machine learning, adeptly manages data distribution differences between various domains. It remains effective when data collection is challenging for the target task, utilizing related data from other domains \cite{liu2019domain}. Further developments in transfer learning and domain adaptation for ChatGPT can affect examining techniques like meta-learning or meta-adaptation to improve its capacity to quickly adapt to new domains. This can implicate learning domain-specific representations or building domain-specific adapters that permit the model to influence its existing knowledge while developing new domain-specific information efficiently. Moreover, research can concentrate on handling the challenge of domain shift, where the distribution of data in the target domain varies significantly from the source domain. Methods such as domain adaptation, domain generalization, or domain robustness can be analyzed to mitigate the impact of domain shift and enhance ChatGPT's performance across diverse domains. Furthermore, improvements in unsupervised or self-supervised learning techniques can enable ChatGPT to leverage unlabeled data from the target domain to enhance domain adaptation. This can concern leveraging techniques like pre-training on large-scale datasets and fine-tuning domain-specific data to fine-tune the model for better performance in the target domain. The investigation of transfer learning and domain adaptation techniques not only helps ChatGPT to adjust to new domains but also decreases the dependence on large quantities of labeled data. This scalability and efficiency are essential for the practical deployment of ChatGPT in real-world applications where training data may be restricted or expensive to acquire. Recent investigation also has been devoted to fine-tuning the ChatGPT model on compact datasets and applying transfer learning to novel challenges. This facilitates enhanced task performance even with limited data availability \cite{jiao2023chatgpt}.

\subsection{Advancing Explainability and Transparency}
In response to the increasing use of AI systems like ChatGPT, there is a growing demand for explainability and transparency. Researchers are actively studying methods to improve the interpretability of ChatGPT's decision-making process, allowing users to acquire insights into how the model develops its responses \cite{hu2023opportunities}. This implicates developing techniques such as rule-based explanations, attention visualization, model introspection, and uncertainty estimation. Enhancing explainability and transparency empowers users with a comprehensive understanding of ChatGPT's underlying reasoning. This promotes trust in the system's outputs and supplies users with greater control over their interactions. By clearing light on the decision-making process, users can assess the reliability of the model's responses and make informed judgments. This promotes a more meaningful and accountable user experience, establishing a foundation for effective collaboration between humans and AI systems like ChatGPT \cite{yang2023evaluations}.

\subsection{Advancing Active Learning and User Feedback}
Future advancements in ChatGPT can focus on developing powerful strategies for dynamic learning and user feedback, improving the model's performance over time. One area of exploration involves query synthesis, where ChatGPT actively generates targeted prompts or questions to request specific feedback from users \cite{neumann2021chatbots}. By managing areas of indecision or ambiguity in the conversation, the model can pursue user input to enhance its understanding and generate more accurate responses. Moreover, researchers can examine adaptive dialogue management strategies that dynamically adapt ChatGPT's responses, conversation style, or conversational purposes based on user feedback \cite{smith2023old}. This adaptive method permits the model to cater to individual user preferences, satisfaction ratings, or clear instructions delivered during the interaction, resulting in more personalized and satisfying conversations. User-initiated teaching can also play a critical role in the model's improvement \cite{gursoy2023chatgpt}. Allowing users to actively teach ChatGPT by delivering direct corrections, offering alternative responses, or sharing specific examples guides the model's behavior. By integrating this feedback, ChatGPT can refine its knowledge and constantly enhance its performance, resulting in more precise and contextually relevant responses \cite{javaid2023study}. To support cumulative learning, ChatGPT can be prepared to update its knowledge incrementally based on user feedback, bypassing the need for large-scale retraining. This permits the model to adjust and improve over time without disrupting its ongoing operation, guaranteeing a seamless user experience \cite{miao2023dao}. Techniques such as reinforcement learning, preference learning, or active sampling can be analyzed to intelligently integrate user feedback into the model's learning process \cite{cooper2023examining}. By including valuable feedback effectively, ChatGPT can make informed updates and refine its conversational capabilities. Moreover, conceiving ChatGPT to deliver explanations or justifications for changes made based on user feedback encourages transparency \cite{shoufan2023exploring}. Users can understand how their feedback influences the model's behavior, promoting trust and enabling more meaningful collaboration between users and ChatGPT. Through the advancement of active learning and user feedback techniques, ChatGPT evolves more adept at learning from and adapting to user interactions \cite{loos2023using}. This iterative feedback loop authorizes users to actively contribute to the model's improvement, resulting in a personalized, refined, and mutually beneficial conversational experience \cite{loos2023using}.

\subsection{Advancing Ethical Considerations}
As AI chatbots like ChatGPT become increasingly popular, addressing ethical considerations is crucial. Future research could focus on ensuring ChatGPT aligns with ethical standards. A pivotal aspect of this development is the preservation of user privacy and data protection \cite{dave2023chatgpt}. Researchers might explore techniques to enhance data anonymization, implement stringent access controls, and limit data retention. To mitigate biased behavior, ChatGPT should undergo thorough bias testing and consistent monitoring \cite{ray2023chatgpt}, addressing biases in the training data and refining the model to ensure fairness across diverse demographics. Transparency and accountability stand as vital ethical pillars \cite{wang2023ethical}. Potential improvements for ChatGPT include offering clear explanations for its responses, highlighting its constraints, and signaling when its knowledge is exhausted. Users must be well-informed about the system's capabilities and constraints, enabling them to make informed decisions and interact responsibly \cite{biswas2023potential}. Regular audits and evaluations can ensure ChatGPT consistently meets ethical benchmarks, with external certifications offering third-party validation of adherence to these standards \cite{li2023advancing}. By emphasizing ethical considerations in ChatGPT's development, AI chatbots can be crafted that prioritize user privacy, fairness, and genuine interaction. Such endeavors build trust, minimize potential harm, and promote the responsible integration of AI technologies in society \cite{zhuo2023exploring}.

\section{Conclusion}
In this comprehensive review, we embarked on a deep exploration of chatbots, elucidating their background, taxonomy, and the plethora of evaluation metrics designed for their assessment. With the exponential rise in technology and the persistent demand for improved human-computer interactions, it has become essential to understand the various methodologies that propel chatbot functionality. From rule-based chatbots, that rely on predefined rules, to the more sophisticated generative models employing neural networks, the landscape of chatbot approaches is both vast and evolving.

Notably, the integration of ChatGPT into FAQ chatbots has emerged as a significant stride forward, harnessing advanced Natural Language Processing techniques and offering unparalleled advantages. These advantages are further amplified when coupled with intent classification, sentiment analysis, and seamless integration with existing knowledge bases.

Nevertheless, like all technological advancements, chatbots come with their share of limitations and ethical quandaries. While their applications span diverse sectors like online shopping, healthcare, and education, they are not exempt from challenges. Yet, these challenges present a multitude of open-ended research opportunities. From the need to optimize conversational flows, to the pursuit of enhancing dialogue across multiple conversational turns and the quest for heightened ethical considerations, the future of chatbots promises a blend of challenges and advancements.

In conclusion, as chatbots continue to integrate deeper into our digital lives, it is paramount to embrace a holistic understanding, critique their current capabilities, and remain curious about their future trajectories. This review serves as a foundation and guidepost for both current practitioners and future innovators in the dynamic realm of chatbot research and development.


\begin{thebibliography}{100}
\expandafter\ifx\csname url\endcsname\relax
  \def\url#1{\texttt{#1}}\fi
\expandafter\ifx\csname urlprefix\endcsname\relax\def\urlprefix{URL }\fi
\expandafter\ifx\csname href\endcsname\relax
  \def\href#1#2{#2} \def\path#1{#1}\fi

\bibitem{kheddar2023deep}
H.~Kheddar, Y.~Himeur, S.~Al-Maadeed, A.~Amira, F.~Bensaali, Deep transfer
  learning for automatic speech recognition: Towards better generalization,
  Knowledge-Based Systems (2023) 110851.

\bibitem{sohail2023decoding}
S.~S. Sohail, F.~Farhat, Y.~Himeur, M.~Nadeem, D.~Øivind Madsen, Y.~Singh,
  S.~Atalla, W.~Mansoor, Decoding chatgpt: A taxonomy of existing research,
  current challenges, and possible future directions, Journal of King Saud
  University - Computer and Information Sciences (2023) 101675.

\bibitem{farhat2023analyzing}
F.~Farhat, E.~S. Silva, H.~Hassani, D.~{\O}. Madsen, S.~S. Sohail, Y.~Himeur,
  M.~A. Alam, A.~Zafar, Analyzing the scholarly footprint of chatgpt: Mapping
  the progress and identifying future trends (2023).

\bibitem{sohail2023future}
S.~S. Sohail, F.~Farhat, Y.~Himeur, M.~Nadeem, D.~{\O}. Madsen, Y.~Singh,
  S.~Atalla, W.~Mansoor, The future of gpt: A taxonomy of existing chatgpt
  research, current challenges, and possible future directions, Current
  Challenges, and Possible Future Directions (April 8, 2023) (2023).

\bibitem{sohail2023using}
S.~S. Sohail, D.~{\O}. Madsen, Y.~Himeur, M.~Ashraf, Using chatgpt to navigate
  ambivalent and contradictory research findings on artificial intelligence,
  Available at SSRN 4413913 (2023).

\bibitem{waisberg2023google}
E.~Waisberg, J.~Ong, M.~Masalkhi, N.~Zaman, P.~Sarker, A.~G. Lee, A.~Tavakkoli,
  Google’s ai chatbot “bard”: a side-by-side comparison with chatgpt and
  its utilization in ophthalmology, Eye (2023) 1--4.

\bibitem{lin2023review}
C.-C. Lin, A.~Y. Huang, S.~J. Yang, A review of ai-driven conversational
  chatbots implementation methodologies and challenges (1999--2022),
  Sustainability 15~(5) (2023) 4012.

\bibitem{jesus2023conversational}
M.~Jes{\'u}s Rodr{\'\i}guez-S{\'a}nchez, K.~Benghazi, D.~Griol, Z.~Callejas,
  Conversational human--machine interfaces, Handbook of Human-Machine Systems
  (2023) 115--123.

\bibitem{sethi2020faq}
F.~Sethi, Faq (frequently asked questions) chatbot for conversation,
  International Journal of Computer Sciences and Engineering 8~(10) (2020).

\bibitem{biswas2023role}
S.~S. Biswas, Role of chat gpt in public health, Annals of Biomedical
  Engineering 51~(5) (2023) 868--869.

\bibitem{lund2023chatting}
B.~D. Lund, T.~Wang, Chatting about chatgpt: how may ai and gpt impact academia
  and libraries?, Library Hi Tech News 40~(3) (2023) 26--29.

\bibitem{king2023conversation}
M.~R. King, ChatGPT, A conversation on artificial intelligence, chatbots, and
  plagiarism in higher education, Cellular and Molecular Bioengineering 16~(1)
  (2023) 1--2.

\bibitem{zhao2023comparing}
X.~Zhao, L.~Chen, Y.~Jin, X.~Zhang, Comparing button-based chatbots with
  webpages for presenting fact-checking results: A case study of health
  information, Information Processing \& Management 60~(2) (2023) 103203.

\bibitem{kitchenham2004procedures}
B.~Kitchenham, Procedures for performing systematic reviews, Keele, UK, Keele
  University 33~(2004) (2004) 1--26.

\bibitem{himeur2021survey}
Y.~Himeur, A.~Alsalemi, A.~Al-Kababji, F.~Bensaali, A.~Amira, C.~Sardianos,
  G.~Dimitrakopoulos, I.~Varlamis, A survey of recommender systems for energy
  efficiency in buildings: Principles, challenges and prospects, Information
  Fusion 72 (2021) 1--21.

\bibitem{chao2021emerging}
M.-H. Chao, A.~J. Trappey, C.-T. Wu, Emerging technologies of natural
  language-enabled chatbots: a review and trend forecast using intelligent
  ontology extraction and patent analytics, Complexity 2021 (2021) 1--26.

\bibitem{abushawar2015alice}
B.~AbuShawar, E.~Atwell, Alice chatbot: Trials and outputs, Computaci{\'o}n y
  Sistemas 19~(4) (2015) 625--632.

\bibitem{mnasri2019recent}
M.~Mnasri, Recent advances in conversational nlp: Towards the standardization
  of chatbot building, arXiv preprint arXiv:1903.09025 (2019).

\bibitem{caldarini2022literature}
G.~Caldarini, S.~Jaf, K.~McGarry, A literature survey of recent advances in
  chatbots, Information 13~(1) (2022) 41.

\bibitem{nithuna2020review}
S.~Nithuna, C.~Laseena, Review on implementation techniques of chatbot, in:
  2020 International Conference on Communication and Signal Processing (ICCSP),
  IEEE, 2020, pp. 0157--0161.

\bibitem{adiwardana2020towards}
D.~Adiwardana, M.-T. Luong, D.~R. So, J.~Hall, N.~Fiedel, R.~Thoppilan,
  Z.~Yang, A.~Kulshreshtha, G.~Nemade, Y.~Lu, et~al., Towards a human-like
  open-domain chatbot, arXiv preprint arXiv:2001.09977 (2020).

\bibitem{liu2016not}
C.-W. Liu, R.~Lowe, I.~V. Serban, M.~Noseworthy, L.~Charlin, J.~Pineau, How not
  to evaluate your dialogue system: An empirical study of unsupervised
  evaluation metrics for dialogue response generation, arXiv preprint
  arXiv:1603.08023 (2016).

\bibitem{shawar2007different}
B.~A. Shawar, E.~Atwell, Different measurement metrics to evaluate a chatbot
  system, in: Proceedings of the workshop on bridging the gap: Academic and
  industrial research in dialog technologies, 2007, pp. 89--96.

\bibitem{mathur2020tangled}
N.~Mathur, T.~Baldwin, T.~Cohn, Tangled up in bleu: Reevaluating the evaluation
  of automatic machine translation evaluation metrics, arXiv preprint
  arXiv:2006.06264 (2020).

\bibitem{rosario2023grading}
G.~Rosario, D.~Noever, Grading conversational responses of chatbots, arXiv
  preprint arXiv:2303.12038 (2023).

\bibitem{khayrallah2023choose}
H.~Khayrallah, Z.~Akhtar, E.~Cohen, J.~Sedoc, How to choose how to choose your
  chatbot: A massively multi-system multireference data set for dialog metric
  evaluation, arXiv preprint arXiv:2305.14533 (2023).

\bibitem{peras2018chatbot}
D.~Peras, Chatbot evaluation metrics, Economic and Social Development: Book of
  Proceedings (2018) 89--97.

\bibitem{mujeeb2017aquabot}
S.~Mujeeb, M.~H. Javed, T.~Arshad, Aquabot: a diagnostic chatbot for
  achluophobia and autism, International Journal of Advanced Computer Science
  and Applications 8~(9) (2017) 209--216.

\bibitem{saha2022healthcare}
B.~Saha, D.~G.~R. Devi, H.~Banerjee, Healthcare chatbot using decision tree
  algorithm, Available at SSRN 4247821 (2022).

\bibitem{mellado2020learning}
R.~Mellado-Silva, A.~Fa{\'u}ndez-Ugalde, M.~B. Lobos, Learning tax regulations
  through rules-based chatbots using decision trees: a case study at the time
  of covid-19, in: 2020 39th International Conference of the Chilean computer
  science society (SCCC), IEEE, 2020, pp. 1--8.

\bibitem{tsai2019ask}
M.-H. Tsai, J.~Y. Chen, S.-C. Kang, Ask diana: A keyword-based chatbot system
  for water-related disaster management, Water 11~(2) (2019) 234.

\bibitem{adamopoulou2020overview}
E.~Adamopoulou, L.~Moussiades, An overview of chatbot technology, in: IFIP
  International Conference on Artificial Intelligence Applications and
  Innovations, Springer, 2020, pp. 373--383.

\bibitem{shawar2002comparison}
B.~A. Shawar, E.~Atwell, A comparison between Alice and Elizabeth chatbot
  systems, University of Leeds, School of Computing research report 2002.19,
  2002.

\bibitem{yamaguchi2018chatbot}
H.~Yamaguchi, M.~Mozgovoy, A.~Danielewicz-Betz, A chatbot based on aiml rules
  extracted from twitter dialogues., in: FedCSIS (Communication Papers), 2018,
  pp. 37--42.

\bibitem{wijaya2020chatbot}
Y.~Wijaya, F.~Zoromi, et~al., Chatbot designing information service for new
  student registration based on aiml and machine learning, JAIA-Journal of
  Artificial Intelligence and Applications 1~(1) (2020) 01--10.

\bibitem{ranoliya2017chatbot}
B.~R. Ranoliya, N.~Raghuwanshi, S.~Singh, Chatbot for university related faqs,
  in: 2017 International Conference on Advances in Computing, Communications
  and Informatics (ICACCI), IEEE, 2017, pp. 1525--1530.

\bibitem{khin2020university}
N.~N. Khin, K.~M. Soe, University chatbot using artificial intelligence markup
  language, in: 2020 IEEE Conference on Computer Applications (ICCA), IEEE,
  2020, pp. 1--5.

\bibitem{buhrke2021making}
J.~B{\"u}hrke, A.~B. Brendel, S.~Lichtenberg, M.~Greve, M.~Mirbabaie, Is making
  mistakes human? on the perception of typing errors in chatbot communication,
  in: Proceedings of the 54th Hawaii International Conference on System
  Sciences, 2021, p. 4456.

\bibitem{thomas2016business}
N.~Thomas, An e-business chatbot using aiml and lsa, in: 2016 International
  conference on advances in computing, communications and informatics (ICACCI),
  IEEE, 2016, pp. 2740--2742.

\bibitem{mikic2018using}
F.~A. Mikic-Fonte, M.~Llamas-Nistal, M.~Caeiro-Rodr{\'\i}guez, Using a
  chatterbot as a faq assistant in a course about computers architecture, in:
  2018 IEEE Frontiers in Education Conference (FIE), IEEE, 2018, pp. 1--4.

\bibitem{li2020conversation}
C.-H. Li, S.-F. Yeh, T.-J. Chang, M.-H. Tsai, K.~Chen, Y.-J. Chang, A
  conversation analysis of non-progress and coping strategies with a banking
  task-oriented chatbot, in: Proceedings of the 2020 CHI Conference on Human
  Factors in Computing Systems, 2020, pp. 1--12.

\bibitem{serban2016building}
I.~Serban, A.~Sordoni, Y.~Bengio, A.~Courville, J.~Pineau, Building end-to-end
  dialogue systems using generative hierarchical neural network models, in:
  Proceedings of the AAAI Conference on Artificial Intelligence, Vol.~30, 2016.

\bibitem{kapovciute2020domain}
J.~Kapo{\v{c}}i{\=u}t{\.e}-Dzikien{\.e}, A domain-specific generative chatbot
  trained from little data, Applied Sciences 10~(7) (2020) 2221.

\bibitem{motger2022software}
Q.~Motger, X.~Franch, J.~Marco, Software-based dialogue systems: survey,
  taxonomy, and challenges, ACM Computing Surveys 55~(5) (2022) 1--42.

\bibitem{serban2017hierarchical}
I.~Serban, A.~Sordoni, R.~Lowe, L.~Charlin, J.~Pineau, A.~Courville, Y.~Bengio,
  A hierarchical latent variable encoder-decoder model for generating
  dialogues, in: Proceedings of the AAAI Conference on Artificial Intelligence,
  Vol.~31, 2017.

\bibitem{xu2017new}
A.~Xu, Z.~Liu, Y.~Guo, V.~Sinha, R.~Akkiraju, A new chatbot for customer
  service on social media, in: Proceedings of the 2017 CHI conference on human
  factors in computing systems, 2017, pp. 3506--3510.

\bibitem{huddar2020dexter}
A.~Huddar, C.~Bysani, C.~Suchak, U.~D. Kolekar, K.~Upadhyaya, Dexter the
  college faq chatbot, in: 2020 International Conference on Convergence to
  Digital World-Quo Vadis (ICCDW), IEEE, 2020, pp. 1--5.

\bibitem{kamphaug2018towards}
{\AA}.~Kamphaug, O.-C. Granmo, M.~Goodwin, V.~I. Zadorozhny, Towards open
  domain chatbots—a gru architecture for data driven conversations, in:
  Internet Science: INSCI 2017 International Workshops, IFIN, DATA ECONOMY,
  DSI, and CONVERSATIONS, Thessaloniki, Greece, November 22, 2017, Revised
  Selected Papers 4, Springer, 2018, pp. 213--222.

\bibitem{liu2019learning}
B.~Liu, M.~Zhao, D.~Niu, K.~Lai, Y.~He, H.~Wei, Y.~Xu, Learning to generate
  questions by learningwhat not to generate, in: The world wide web conference,
  2019, pp. 1106--1118.

\bibitem{wang2019answer}
W.~Wang, S.~Feng, D.~Wang, Y.~Zhang, Answer-guided and semantic coherent
  question generation in open-domain conversation, in: Proceedings of the 2019
  Conference on Empirical Methods in Natural Language Processing and the 9th
  International Joint Conference on Natural Language Processing (EMNLP-IJCNLP),
  2019, pp. 5066--5076.

\bibitem{egonmwan2019transformer}
E.~Egonmwan, Y.~Chali, Transformer and seq2seq model for paraphrase generation,
  in: Proceedings of the 3rd Workshop on Neural Generation and Translation,
  2019, pp. 249--255.

\bibitem{boussakssou2022chatbot}
M.~Boussakssou, H.~Ezzikouri, M.~Erritali, Chatbot in arabic language using seq
  to seq model, Multimedia Tools and Applications 81~(2) (2022) 2859--2871.

\bibitem{khin2020question}
N.~N. Khin, K.~M. Soe, Question answering based university chatbot using
  sequence to sequence model, in: 2020 23rd Conference of the Oriental COCOSDA
  International Committee for the Co-ordination and Standardisation of Speech
  Databases and Assessment Techniques (O-COCOSDA), IEEE, 2020, pp. 55--59.

\bibitem{ricciardelli2019self}
E.~Ricciardelli, D.~Biswas, Self-improving chatbots based on reinforcement
  learning, in: 4th Multidisciplinary Conference on Reinforcement Learning and
  Decision Making, 2019.

\bibitem{sassoon2019storytelling}
J.~Sassoon, Storytelling e intelligenza artificiale, Quando le storie le
  raccontano i robot. FrancoAngeli, Milano (2019).

\bibitem{bailey2021digital}
D.~Bailey, A.~Southam, J.~Costley, Digital storytelling with chatbots: Mapping
  l2 participation and perception patterns, Interactive Technology and Smart
  Education 18~(1) (2021) 85--103.

\bibitem{serban2017deep}
I.~V. Serban, C.~Sankar, M.~Germain, S.~Zhang, Z.~Lin, S.~Subramanian, T.~Kim,
  M.~Pieper, S.~Chandar, N.~R. Ke, et~al., A deep reinforcement learning
  chatbot, arXiv preprint arXiv:1709.02349 (2017).

\bibitem{xu2022elinor}
Y.~Xu, V.~Vigil, A.~S. Bustamante, M.~Warschauer, “elinor’s talking to
  me!”: Integrating conversational ai into children’s narrative science
  programming, in: Proceedings of the 2022 CHI Conference on Human Factors in
  Computing Systems, 2022, pp. 1--16.

\bibitem{nugraha2022chatbot}
M.~A. Nugraha, Z.~Baizal, D.~Richasdy, Chatbot-based movie recommender system
  using pos tagging, Building of Informatics, Technology and Science (BITS)
  4~(2) (2022) 624--630.

\bibitem{bao2020plato}
S.~Bao, H.~He, F.~Wang, H.~Wu, H.~Wang, W.~Wu, Z.~Guo, Z.~Liu, X.~Xu, Plato-2:
  Towards building an open-domain chatbot via curriculum learning, arXiv
  preprint arXiv:2006.16779 (2020).

\bibitem{raj2023generative}
V.~Raj, M.~Phridviraj, A generative model based chatbot using recurrent neural
  networks, in: Advanced Network Technologies and Intelligent Computing: Second
  International Conference, ANTIC 2022, Varanasi, India, December 22--24, 2022,
  Proceedings, Part II, Springer, 2023, pp. 379--392.

\bibitem{lhasiw2021bidirectional}
N.~Lhasiw, N.~Sanglerdsinlapachai, T.~Tanantong, A bidirectional lstm model for
  classifying chatbot messages, in: 2021 16th International Joint Symposium on
  Artificial Intelligence and Natural Language Processing (iSAI-NLP), IEEE,
  2021, pp. 1--6.

\bibitem{muangkammuen2018automated}
P.~Muangkammuen, N.~Intiruk, K.~R. Saikaew, Automated thai-faq chatbot using
  rnn-lstm, in: 2018 22nd International Computer Science and Engineering
  Conference (ICSEC), IEEE, 2018, pp. 1--4.

\bibitem{da2019ibm}
J.~da~Silva~Oliveira, D.~B. Esp{\'\i}ndola, R.~Barwaldt, L.~M. Ribeiro,
  M.~Pias, Ibm watson application as faq assistant about moodle, in: 2019 ieee
  frontiers in education conference (FIE), IEEE, 2019, pp. 1--8.

\bibitem{vu2021online}
T.~L. Vu, K.~Z. Tun, C.~Eng-Siong, R.~E. Banchs, Online faq chatbot for
  customer support, in: Increasing Naturalness and Flexibility in Spoken
  Dialogue Interaction: 10th International Workshop on Spoken Dialogue Systems,
  Springer, 2021, pp. 251--259.

\bibitem{reyes2019methodology}
R.~Reyes, D.~Garza, L.~Garrido, V.~De~la Cueva, J.~Ramirez, Methodology for the
  implementation of virtual assistants for education using google dialogflow,
  in: Advances in Soft Computing: 18th Mexican International Conference on
  Artificial Intelligence, MICAI 2019, Xalapa, Mexico, October 27--November 2,
  2019, Proceedings 18, Springer, 2019, pp. 440--451.

\bibitem{hulman2023chatgpt}
A.~Hulman, O.~L. Dollerup, J.~F. Mortensen, M.~Fenech, K.~Norman, H.~Stoevring,
  T.~K. Hansen, Chatgpt-versus human-generated answers to frequently asked
  questions about diabetes: a turing test-inspired survey among employees of a
  danish diabetes center, medRxiv (2023) 2023--02.

\bibitem{dwivedi2023so}
Y.~K. Dwivedi, N.~Kshetri, L.~Hughes, E.~L. Slade, A.~Jeyaraj, A.~K. Kar, A.~M.
  Baabdullah, A.~Koohang, V.~Raghavan, M.~Ahuja, et~al., “so what if chatgpt
  wrote it?” multidisciplinary perspectives on opportunities, challenges and
  implications of generative conversational ai for research, practice and
  policy, International Journal of Information Management 71 (2023) 102642.

\bibitem{sun2018overview}
P.~Sun, X.~Yang, X.~Zhao, Z.~Wang, An overview of named entity recognition, in:
  2018 International Conference on Asian Language Processing (IALP), IEEE,
  2018, pp. 273--278.

\bibitem{hu2023zero}
Y.~Hu, I.~Ameer, X.~Zuo, X.~Peng, Y.~Zhou, Z.~Li, Y.~Li, J.~Li, X.~Jiang,
  H.~Xu, Zero-shot clinical entity recognition using chatgpt, arXiv preprint
  arXiv:2303.16416 (2023).

\bibitem{cegin2023chatgpt}
J.~Cegin, J.~Simko, P.~Brusilovsky, Chatgpt to replace crowdsourcing of
  paraphrases for intent classification: Higher diversity and comparable model
  robustness, arXiv preprint arXiv:2305.12947 (2023).

\bibitem{essop2023developing}
L.~Essop, A.~Singh, J.~Wing, Developing a comprehensive evaluation
  questionnaire for university faq administration chatbots, in: 2023 Conference
  on Information Communications Technology and Society (ICTAS), IEEE, 2023, pp.
  1--7.

\bibitem{alemdag2023effect}
E.~Alemdag, The effect of chatbots on learning: a meta-analysis of empirical
  research, Journal of Research on Technology in Education (2023) 1--23.

\bibitem{ge2023designing}
Q.~Ge, L.~Liu, H.~Zhang, L.~Li, X.~Li, X.~Zhu, L.~Liao, D.~Song, Designing
  philobot: A chatbot for mental health support with cbt techniques, in:
  Chinese Intelligent Automation Conference, Springer, 2023, pp. 361--371.

\bibitem{ait2023impact}
T.~Ait~Baha, M.~El~Hajji, Y.~Es-Saady, H.~Fadili, The impact of educational
  chatbot on student learning experience, Education and Information
  Technologies (2023) 1--24.

\bibitem{balderas2023chatbot}
A.~Balderas, R.~F. Garc{\'\i}a-Mena, M.~Huerta, N.~Mora, J.~M. Dodero, Chatbot
  for communicating with university students in emergency situation, Heliyon
  9~(9) (2023).

\bibitem{chen2023adoption}
T.~Chen, M.~Gasc{\'o}-Hernandez, M.~Esteve, The adoption and implementation of
  artificial intelligence chatbots in public organizations: Evidence from us
  state governments, The American Review of Public Administration (2023)
  02750740231200522.

\bibitem{omar2023chatgpt}
R.~Omar, O.~Mangukiya, P.~Kalnis, E.~Mansour, Chatgpt versus traditional
  question answering for knowledge graphs: Current status and future directions
  towards knowledge graph chatbots, arXiv preprint arXiv:2302.06466 (2023).

\bibitem{hariri2023unlocking}
W.~Hariri, Unlocking the potential of chatgpt: A comprehensive exploration of
  its applications, advantages, limitations, and future directions in natural
  language processing, arXiv preprint arXiv:2304.02017 (2023).

\bibitem{limna2023use}
P.~Limna, T.~Kraiwanit, K.~Jangjarat, P.~Klayklung, P.~Chocksathaporn, The use
  of chatgpt in the digital era: Perspectives on chatbot implementation,
  Journal of Applied Learning and Teaching 6~(1) (2023).

\bibitem{sallam2023chatgpt}
M.~Sallam, N.~Salim, M.~Barakat, A.~Al-Tammemi, Chatgpt applications in
  medical, dental, pharmacy, and public health education: A descriptive study
  highlighting the advantages and limitations, Narra J 3~(1) (2023) e103--e103.

\bibitem{ray2023chatgpt}
P.~P. Ray, Chatgpt: A comprehensive review on background, applications, key
  challenges, bias, ethics, limitations and future scope, Internet of Things
  and Cyber-Physical Systems (2023).

\bibitem{george2023review}
A.~S. George, A.~H. George, A review of chatgpt ai's impact on several business
  sectors, Partners Universal International Innovation Journal 1~(1) (2023)
  9--23.

\bibitem{paul2023chatgpt}
J.~Paul, A.~Ueno, C.~Dennis, Chatgpt and consumers: Benefits, pitfalls and
  future research agenda (2023).

\bibitem{li2023chatgpt}
J.~Li, A.~Dada, J.~Kleesiek, J.~Egger, Chatgpt in healthcare: A taxonomy and
  systematic review, medRxiv (2023) 2023--03.

\bibitem{cascella2023evaluating}
M.~Cascella, J.~Montomoli, V.~Bellini, E.~Bignami, Evaluating the feasibility
  of chatgpt in healthcare: an analysis of multiple clinical and research
  scenarios, Journal of Medical Systems 47~(1) (2023) 33.

\bibitem{asch2023interview}
D.~A. Asch, An interview with chatgpt about health care, NEJM Catalyst
  Innovations in Care Delivery 4~(2) (2023).

\bibitem{javaid2023chatgpt}
M.~Javaid, A.~Haleem, R.~P. Singh, Chatgpt for healthcare services: An emerging
  stage for an innovative perspective, BenchCouncil Transactions on Benchmarks,
  Standards and Evaluations 3~(1) (2023) 100105.

\bibitem{tlili2023if}
A.~Tlili, B.~Shehata, M.~A. Adarkwah, A.~Bozkurt, D.~T. Hickey, R.~Huang,
  B.~Agyemang, What if the devil is my guardian angel: Chatgpt as a case study
  of using chatbots in education, Smart Learning Environments 10~(1) (2023) 15.

\bibitem{sok2023chatgpt}
S.~Sok, K.~Heng, Chatgpt for education and research: A review of benefits and
  risks, Available at SSRN 4378735 (2023).

\bibitem{opara2023chatgpt}
E.~Opara, A.~Mfon-Ette~Theresa, T.~C. Aduke, Chatgpt for teaching, learning and
  research: Prospects and challenges, Opara Emmanuel Chinonso, Adalikwu
  Mfon-Ette Theresa, Tolorunleke Caroline Aduke (2023). ChatGPT for Teaching,
  Learning and Research: Prospects and Challenges. Glob Acad J Humanit Soc Sci
  5 (2023).

\bibitem{ayanouz2020smart}
S.~Ayanouz, B.~A. Abdelhakim, M.~Benhmed, A smart chatbot architecture based
  nlp and machine learning for health care assistance, in: Proceedings of the
  3rd International Conference on Networking, Information Systems \& Security,
  2020, pp. 1--6.

\bibitem{zhou2020design}
L.~Zhou, J.~Gao, D.~Li, H.-Y. Shum, The design and implementation of xiaoice,
  an empathetic social chatbot, Computational Linguistics 46~(1) (2020) 53--93.

\bibitem{chhabra2021overview}
A.~Chhabra, K.~Masalkovait{\.e}, P.~Mohapatra, An overview of fairness in
  clustering, IEEE Access 9 (2021) 130698--130720.

\bibitem{singh2023chatgpt}
D.~Singh, Chatgpt: A new approach to revolutionise organisations, ugc approved
  research journals in india| UGC Newly Added Journals|(IJNMS) 10~(1) (2023)
  57--63.

\bibitem{singh2023leveraging}
G.~Singh, Leveraging chatgpt for real-time decision-making in autonomous
  systems, Eduzone: International Peer Reviewed/Refereed Multidisciplinary
  Journal 12~(2) (2023) 101--106.

\bibitem{lum2022can}
Z.~C. Lum, Can artificial intelligence pass the american board of orthopaedic
  surgery examination? orthopaedic residents versus chatgpt, Clinical
  Orthopaedics and Related Research{\textregistered} (2022) 10--1097.

\bibitem{loni2011survey}
B.~Loni, A survey of state-of-the-art methods on question classification, Delft
  University of Technology, Tech. Rep 55 (2011) 57.

\bibitem{liu2019domain}
Y.~Liu, C.~Yang, K.~Liu, B.~Chen, Y.~Yao, Domain adaptation transfer learning
  soft sensor for product quality prediction, Chemometrics and Intelligent
  Laboratory Systems 192 (2019) 103813.

\bibitem{jiao2023chatgpt}
W.~Jiao, W.~Wang, J.-t. Huang, X.~Wang, Z.~Tu, Is chatgpt a good translator? a
  preliminary study, arXiv preprint arXiv:2301.08745 (2023).

\bibitem{hu2023opportunities}
X.~Hu, Y.~Tian, K.~Nagato, M.~Nakao, A.~Liu, Opportunities and challenges of
  chatgpt for design knowledge management, arXiv preprint arXiv:2304.02796
  (2023).

\bibitem{yang2023evaluations}
K.~Yang, S.~Ji, T.~Zhang, Q.~Xie, S.~Ananiadou, On the evaluations of chatgpt
  and emotion-enhanced prompting for mental health analysis, arXiv preprint
  arXiv:2304.03347 (2023).

\bibitem{neumann2021chatbots}
A.~T. Neumann, T.~Arndt, L.~K{\"o}bis, R.~Meissner, A.~Martin, P.~de~Lange,
  N.~Pengel, R.~Klamma, H.-W. Wollersheim, Chatbots as a tool to scale
  mentoring processes: Individually supporting self-study in higher education,
  Frontiers in artificial intelligence 4 (2021) 668220.

\bibitem{smith2023old}
A.~Smith, S.~Hachen, R.~Schleifer, D.~Bhugra, A.~Buadze, M.~Liebrenz, Old dog,
  new tricks? exploring the potential functionalities of chatgpt in supporting
  educational methods in social psychiatry, International Journal of Social
  Psychiatry (2023) 00207640231178451.

\bibitem{gursoy2023chatgpt}
D.~Gursoy, Y.~Li, H.~Song, Chatgpt and the hospitality and tourism industry: an
  overview of current trends and future research directions, Journal of
  Hospitality Marketing \& Management 32~(5) (2023) 579--592.

\bibitem{javaid2023study}
M.~Javaid, A.~Haleem, R.~P. Singh, A study on chatgpt for industry 4.0:
  Background, potentials, challenges, and eventualities, Journal of Economy and
  Technology (2023).

\bibitem{miao2023dao}
Q.~Miao, W.~Zheng, Y.~Lv, M.~Huang, W.~Ding, F.-Y. Wang, Dao to hanoi via
  desci: Ai paradigm shifts from alphago to chatgpt, IEEE/CAA Journal of
  Automatica Sinica 10~(4) (2023) 877--897.

\bibitem{cooper2023examining}
G.~Cooper, Examining science education in chatgpt: An exploratory study of
  generative artificial intelligence, Journal of Science Education and
  Technology 32~(3) (2023) 444--452.

\bibitem{shoufan2023exploring}
A.~Shoufan, Exploring students’ perceptions of chatgpt: Thematic analysis and
  follow-up survey, IEEE Access (2023).

\bibitem{loos2023using}
E.~Loos, J.~Gr{\"o}pler, M.-L. S.~S. Goudeau, Using chatgpt in education: Human
  reflection on gpt’s self-reflection, Societies 13~(8) (2023) 196.

\bibitem{dave2023chatgpt}
T.~Dave, S.~A. Athaluri, S.~Singh, Chatgpt in medicine: an overview of its
  applications, advantages, limitations, future prospects, and ethical
  considerations, Frontiers in Artificial Intelligence 6 (2023) 1169595.

\bibitem{wang2023ethical}
C.~Wang, S.~Liu, H.~Yang, J.~Guo, Y.~Wu, J.~Liu, Ethical considerations of
  using chatgpt in health care, Journal of Medical Internet Research 25 (2023)
  e48009.

\bibitem{biswas2023potential}
S.~S. Biswas, Potential use of chat gpt in global warming, Annals of biomedical
  engineering 51~(6) (2023) 1126--1127.

\bibitem{li2023advancing}
S.~Li, Z.~Guo, X.~Zang, Advancing the production of clinical medical devices
  through chatgpt, Annals of Biomedical Engineering (2023) 1--5.

\bibitem{zhuo2023exploring}
T.~Y. Zhuo, Y.~Huang, C.~Chen, Z.~Xing, Exploring ai ethics of chatgpt: A
  diagnostic analysis, arXiv preprint arXiv:2301.12867 (2023).

\end{thebibliography}
\end{document}